\documentclass[twocolumn, preprintnumbers,
endnote,nofootinbib,plb,9pt,superscriptaddress]{revtex4}

\usepackage[utf8]{inputenc}
\usepackage{graphicx}
\usepackage{amssymb}
\usepackage{amsxtra}
\usepackage{amsmath}
\usepackage{booktabs,multirow,tabularx}
\usepackage{braket}
\usepackage{slashed}
\usepackage{float}
\usepackage{placeins}
\usepackage{rotating}
\usepackage{lscape}
\usepackage{color}
\usepackage{hyperref}
\usepackage{enumitem}

\usepackage[Q=yes,pverb-linebreak=no]{examplep}

\DeclareGraphicsExtensions{.eps}
\graphicspath{{./}}

\newcommand{\be}{\begin{equation}}
\newcommand{\ee}{\end{equation}}

\def\beq{\begin{equation}}
\def\bea{\begin{eqnarray}}
\def\eeq{\end{equation}}
\def\eea{\end{eqnarray}}
\def\beqnl{\begin{align}}
\def\endal{\end{align}}

\makeatletter
\newcommand{\normalorbold}{%
  \ifnum\pdf@strcmp{\math@version}{bold}=\z@ bx\else m\fi
}
\makeatother

\begin{document}

\title{\boldmath Top quark mass effects in $gg\to ZZ$ at two loops and off-shell Higgs interference}

\author{Ramona Gr\"{o}ber\footnote{Email: ramona.groeber@physik.hu-berlin.de}
}
\affiliation{Institut f\"ur Physik, Humboldt-Universit\"at zu Berlin, 12489 Berlin, Germany}

\author{Andreas Maier\footnote{Email: andreas.martin.maier@desy.de}
}
\affiliation{Deutsches Elektronen–Synchrotron, DESY, Platanenallee 6, D–15738 Zeuthen, Germany}
\author{Thomas Rauh\footnote{Email: rauh@itp.unibe.ch}
}
\affiliation{Albert Einstein Center for Fundamental Physics,
Institute for Theoretical Physics, University of Bern,
Sidlerstrasse 5, CH-3012 Bern, Switzerland}
{\widetext{
\begin{flushright}
  DESY 19-136\\
  HU-EP-19/23\\
  SAGEX-19-18
\end{flushright}
         }

}

\begin{abstract}
We consider top-quark mass effects in the Higgs-interference
contribution to $Z$-boson pair production in gluon fusion. While this
production mechanism is formally of next-to-next-to leading order, its
contribution is numerically important above the top threshold
$M_{ZZ}^2=4m_t^2$.  This region is essential to constrain the width of
the Higgs boson and good control over the top-quark mass dependence is
crucial. We determine the form factors that are relevant for the
interference contribution at two-loop order using a method based on a
conformal mapping and Pad\'e approximants constructed from the
expansions of the amplitude for large top mass and around the top
threshold.
\end{abstract}

\maketitle


\section{Introduction\label{sec:intro}}

A direct measurement of the Higgs boson width $\Gamma_H$ is not possible at the LHC or even          
the envisioned next generation of collider experiments.
However, indirect constraints
can be obtained at the LHC by studying the process $pp\to H\to ZZ\,(\to4l)$ on the
Higgs boson peak where the cross section depends on the combination $g_{Hgg}^2g_{HZZ}^2/\Gamma_H$
and off the peak where the measurement of the cross section constrains
the product $g_{Hgg}^2g_{HZZ}^2$ of the effective Higgs boson-gluon
coupling $g_{Hgg}$ and the Higgs boson-$Z$ boson coupling $g_{HZZ}$, as proposed
in~\cite{Kauer:2012hd,Caola:2013yja,Campbell:2013una}.\footnote{Note that the indirect 
way of constraining the Higgs width  is not entirely model-independent 
\cite{Englert:2014aca, Englert:2014ffa}.}
The same strategy can be employed with $WW$ final states~\cite{Campbell:2013wga}.
The latest studies from the LHC experiments give an upper limit of 14.4\,MeV at 95\%
C.L. from the $ZZ$ final state at ATLAS~\cite{Aaboud:2018puo} and the value $3.2^{+2.8}_{-2.2}$\,MeV
from the combination of $VV$ final states in CMS~\cite{Sirunyan:2019twz}, close to the
SM prediction $\Gamma_H^\text{SM}=4.10\pm0.06$\,MeV~\cite{deFlorian:2016spz}.
Measurements of the Higgs boson signal at large invariant mass can also be used to directly
constrain physics beyond the Standard Model in the Higgs
sector~\cite{Gainer:2014hha,Azatov:2014jga,Buschmann:2014sia,Azatov:2016xik}.

Here, we focus on the loop-induced continuum gluon fusion process $gg\to ZZ$ and in particular
its interference with the off-shell Higgs contribution $gg\to H^*\to ZZ$. Despite the
narrow width of the Higgs boson these interference effects are sizable with 10\% of the Higgs signal
stemming from the off-shell region where the invariant mass of the two decay products is greater 
than $2\,m_Z$~\cite{Kauer:2012hd}
and higher-order corrections are required to control the uncertainties. The Higgs-mediated
amplitude only depends on two scales, the mass $m_q$ of the quark in the loop and the invariant
mass $M_{ZZ}$ of the final state. Next-to-leading order (NLO)
corrections with the full quark-mass dependence have been known for some
time~\cite{Spira:1995rr,Harlander:2005rq,Anastasiou:2006hc,Aglietti:2006tp},
and the top-quark mass dependence at next-to-next-to leading order (NNLO) has been reconstructed very
recently~\cite{Davies:2019nhm} (see also~\cite{Harlander:2019ioe}).
On the other hand the continuum amplitude depends on four scales $m_q$, $m_Z$, $M_{ZZ}$
and the transverse momentum $p_T$ of one of the $Z$ bosons, and the exact result is only known
at leading order (LO)~\cite{Glover:1988rg} while an analytic NLO calculation appears extremely challenging.
In the massless limit $m_q=0$ the two-loop amplitude has been determined
in~\cite{Caola:2015ila,vonManteuffel:2015msa,Caola:2015psa} and the NLO cross section in \cite{Caola:2016trd}.
Recently, also the quark–gluon channel has been included \cite{Grazzini:2018owa}.

The contribution from top quarks at two-loop order has been computed in
a large-mass expansion
(LME)~\cite{Melnikov:2015laa,Campbell:2016ivq,Caola:2016trd} and is
known up to $1/m_t^{12}$.  While the contribution from massless quarks
dominates the interference correction at small invariant masses, the
top-quark contribution is of the same size near the top threshold
$M_{ZZ} = 4m_t^2$ and dominates in the large invariant-mass
regime. Since the LME ceases to provide a reliable description above the
top threshold, the authors of~\cite{Campbell:2016ivq} have improved
their prediction by a conformal mapping and the construction of Pad\'e
approximants based on the available number of LME
coefficients. In~\cite{Grober:2017uho} we have extended this method by
considering the expansion around the top threshold in addition to the
LME and demonstrated that the top-mass effects can be reproduced
correctly by comparing results for the two-loop amplitude for $gg\to HH$
with the numerical calculation
from~\cite{Borowka:2016ehy,Borowka:2016ypz,Heinrich:2017kxx}.\footnote{Recently,
an independent numerical calculation~\cite{Baglio:2018lrj}, several
approximations
\cite{Bonciani:2018omm,Davies:2018ood,Davies:2018qvx,Xu:2018eos} which
are consistent with the earlier results and a combined result
\cite{Davies:2019dfy} have appeared.}

In this work we consider the form factors of the continuum $gg\to ZZ$ amplitude that are
relevant for the interference contribution at one and two loops. The non-analytic terms
in the expansion around the top threshold are computed up to at least order $(1-z)^4$, where
$z=M_{ZZ}^2/(4m_t^2)+i0$, and used to construct Pad\'e approximants. Together with the
exactly known real NLO top quark \cite{Campbell:2014gua,Campbell:2016ivq} and the massless
quark corrections \cite{Caola:2015ila,vonManteuffel:2015msa,Caola:2015psa,Caola:2016trd,Grazzini:2018owa}
this is sufficient to determine the full NLO interference contribution with realistic
top-quark mass dependence.


\section{Form factors for interference\label{sec:amplitude}}

\begin{figure}[h]
  \includegraphics{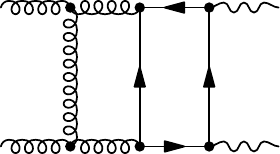}
  \qquad
  \includegraphics{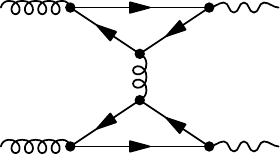}
  \centering
  \caption{Examples for box (left) and double-triangle (right) top-mediated
    contributions to $gg \to ZZ$.}
  \label{fig:box_triangle}
\end{figure}

Up to the two loop level, the amplitude for the top-mediated
non-resonant continuum production process
$g(\mu,A,p_1)+g(\nu,B,p_2)\to Z(\alpha,p_3)+Z(\beta,p_4)$ receives
contributions from both box and double-triangle diagrams, see
figure~\ref{fig:box_triangle}. The latter are known for arbitrary quark
masses~\cite{Campbell:2016ivq, Campbell:2007ev} and will not be
discussed in the following.

The box amplitude $\Ket{B_{\mu\nu\alpha\beta}^{AB}}$ has a complicated tensor
structure~\cite{Glover:1988rg,Costantini:1971cj,Caola:2015ila,vonManteuffel:2015msa}.
However, the interference with the Higgs-mediated amplitude is described by a single form factor.
Adopting the conventions of \cite{Campbell:2016ivq} it takes the form
\begin{equation}
 \Ket{\mathcal{B}} = \frac{\delta^{AB}}{N_A}(p_1\cdot p_2g^{\mu\nu}-p_1^\nu p_2^\mu)
                     P_Z^{\alpha\rho}(p_3)P_{Z,\rho}^\beta(p_4)\Ket{B_{\mu\nu\alpha\beta}^{AB}},
\label{eq:projector}
\end{equation}
with $N_A=N_c^2-1$ and $P_Z^{\alpha\rho}(p) = -g^{\alpha\rho}+p^\alpha
p^\rho/m_Z^2$. The form factor can be decomposed into a vector and axial-vector part
\begin{equation}
 \Ket{\mathcal{B}} = \frac{ig_W^2}{4\cos^2\theta_W}
 \left(v_t^2\Ket{\widetilde{\mathcal{B}}_{VV}} + a_t^2\Ket{\widetilde{\mathcal{B}}_{AA}}\right),
 \label{eq:VVandAA}
\end{equation}
where $a_t=1/2$ and $v_t=1/2-4/3\sin^2\theta_W$ denote the axial-vector and vector couplings
for an up-type quark. Mixed $v_ta_t$ terms are forbidden by charge conjugation symmetry.
The order in the strong coupling constant $\alpha_s$ is indicated as follows
\begin{equation}
 \Ket{\widetilde{\mathcal{B}}_i} = \frac{\alpha_s}{4\pi}\Ket{\widetilde{\mathcal{B}}_i^{(1)}} +
                     \left(\frac{\alpha_s}{4\pi}\right)^2\Ket{\widetilde{\mathcal{B}}_i^{(2)}} + \dots \,,
 \label{eq:alphas_expansion}
\end{equation}
with $i=VV,AA$. At order $\alpha_s^2$ the renormalized form factors contain IR divergences,
which cancel in the combination with real corrections, and we define the finite remainder by
applying the subtraction~\cite{Catani:1998bh}\footnote{Note that
  this subtraction differs at order $\epsilon^0$ from the one given in
  eq. (2.14) of~\cite{Campbell:2016ivq}.}
\begin{equation}
  \label{eq:IR_subtr}
  \Ket{\widetilde{\mathcal{F}}_i^{(2)}} =
    \Ket{\widetilde{\mathcal{B}}_{i}^{(2)}} +
    \frac{e^{\epsilon\gamma_E}}{\Gamma(1-\epsilon)}\left[\frac{2C_A}{\epsilon^2}\left(\frac{\mu^2}{-s}\right)^\epsilon
      + \frac{\beta_0}{\epsilon}\right]\Ket{\widetilde{\mathcal{B}}_{i}^{(1)}},
\end{equation}
where $\beta_0 = \tfrac{11}{3}C_A - \tfrac{4}{3} T_f n_l, C_A = 3, T_f =
\tfrac{1}{2}, n_l = 5$, and the form factors
$\Ket{\widetilde{\mathcal{B}}_{i}^{(1,2)}}$ are defined in
$d=4-2\epsilon$ dimensions. The one-loop form-factors
$\Ket{\widetilde{\mathcal{B}}_{i}^{(1)}}$ are already finite; we define
$\Ket{\widetilde{\mathcal{F}}_{i}^{(1)}} =
\Ket{\widetilde{\mathcal{B}}_{i}^{(1)}}$ for the sake of a consistent notation.

\subsection{The amplitude near threshold\label{sec:threshold}}

Above the top threshold at $z=1$ the top quarks in the loop can go on shell which
manifests as non-analytic terms in the expansion of the form factors in $\bar{z}\equiv1-z$,
generating a sizable imaginary part. As shown in \cite{Grober:2017uho} the knowledge
of these terms alone provides very valuable information for the determination
of top-quark mass effects in our approach. The calculation of the non-analytic terms is
significantly simpler than that of the analytic contributions and was described in detail
in~\cite{Grober:2017uho} for the three leading non-analytic expansion terms of the one
and two-loop form factors for $gg\to HH$. For $gg\to ZZ$ we expand the amplitude up to
high orders in $\bar{z}\equiv1-z$
and therefore use the expansion by regions~\cite{Beneke:1997zp,Jantzen:2011nz} to
expand the full-theory diagrams instead of an EFT approach
where a large number of effective vertices is required due to the deep expansion.
We use \texttt{QGRAF}~\cite{Nogueira:1991ex} to generate the Feynman diagrams which
are processed and expanded using private \texttt{FORM}~\cite{Vermaseren:2000nd} code.
The IBP reduction~\cite{Chetyrkin:1981qh} is performed with \texttt{FIRE}~\cite{Smirnov:2014hma}
which is based on the Laporta algorithm~\cite{Laporta:2001dd}.

Our results are given in Appendix~\ref{sec:F_thr_res} and an ancillary \texttt{Mathematica}
file. They are of the form
\begin{align}
\label{eq:formfactor_expansion}
 \Ket{\widetilde{\mathcal{F}}_i^{(1)}} & \mathop{\asymp}\limits^{z\to1}
    \sum_{n=3}^\infty a_i^{(n,0)}\bar{z}^{\frac{n}{2}}\,,\\
 \Ket{\widetilde{\mathcal{F}}_i^{(2)}} & \mathop{\asymp}\limits^{z\to1}
    \sum_{n=2}^\infty \,\sum_{m=\bar{n}_2}^{1} \left[b_i^{(n,m)} +
    b_{i,\ln}^{(n,m)}\ln(-4z)\right]\bar{z}^{\frac{n}{2}}\ln^m\bar{z}\,,\nonumber
\end{align}
where $\bar{n}_2$ is $n$ modulo 2, the coefficients are functions of the dimensionless
variables $r_Z=m_Z^2/M_{ZZ}^2$ and $\tilde{x}=(p_T^2+m_Z^2)/M_{ZZ}^2$. We use the symbol
$\asymp$ to indicate that terms which are analytic in $\bar{z}$ and currently unknown
have been dropped on the right-hand side.

Threshold logarithms $\ln\bar{z}$ and logarithms $\ln(-4z)$ related to
massless cuts in the amplitude first appear at two-loop order. While we
generally compute the expansion coefficients up to $n=8$, i.e. expand up
to $\bar{z}^4$, we find that for the massless-cut contribution
proportional to $\ln(-4z)$ more input is required to achieve a reliable
Padé approximation. We therefore compute the corresponding coefficients
$b_{i,\ln}^{(n,m)}$ up to $n=9$.

As in Higgs pair production there is no S-wave contribution to the form
factors relevant for the interference and the leading non-analytic terms
involve the $\bar{z}$-suppressed P-wave Green
function~\cite{Beneke:2013kia}.

\subsection{Behavior for $z\to \infty$\label{sec:sme}}

In addition to the LME and threshold expansions we can exploit scaling
information in the small-mass limit $m_t\to0$ which corresponds to
$z\to\infty$. This does not require an additional calculation in this
region but relies solely on the symmetries of QCD. The absence of
infrared $1/m_t$ power divergences as $m_t\to0$ implies that the form
factors can only show logarithmic behavior as $z\to\infty$. Below we
show that the difference
\begin{equation}
\Ket{\widetilde{\mathcal{B}}_{AA-VV}} \equiv
\Ket{\widetilde{\mathcal{B}}_{AA}} - \Ket{\widetilde{\mathcal{B}}_{VV}}
\end{equation}
vanishes as $z\to\infty$. To prove this we note that
chirality is conserved in massless QCD and hence the four-point
correlator of two vector currents, a left-handed and a right-handed
current, which we denote in short by [V,V,V-A,V+A], vanishes in the
limit of zero quark masses.\footnote{To make the double-triangle
contribution shown in Fig.~\ref{fig:box_triangle} anomaly free we have
to consider doublets of quarks and not just a single (top) quark, but we
omit this technicality here since the double-triangle contribution is
known and not considered below.} Using that the correlator [V,V,V,A]
vanishes due to charge conjugation we immediately conclude that
[V,V,V,V]-[V,V,A,A]$\to0$ as $z\to\infty$. We exploit this below and reconstruct
the top-mass dependence of $\Ket{\widetilde{\mathcal{F}}_{VV}^{(i)}}$
and $\Ket{\widetilde{\mathcal{F}}_{AA}^{(i)}}-\Ket{\widetilde{\mathcal{F}}_{VV}^{(i)}}$
where we have one additional condition for the latter.


\section{The method\label{sec:method}}

We approximate the box form factors~\eqref{eq:VVandAA} using our
approach from~\cite{Grober:2017uho}. First, we introduce subtraction
functions $s_{VV}^{(2)}, s_{AA}^{(2)}$ in such a way that the combinations
$\Ket{\widetilde{\mathcal{F}}_i^{(2)}} - s_i^{(2)}$ retain their analytic
structure for $|z| < 1$ but have threshold expansions which are free of
logarithms $\ln(\bar{z})$ up to the highest known order, i.e.~up to
$\bar{z}^4$. The construction of such subtraction functions is detailed
in~\cite{Grober:2017uho} and we give the ones we explicitly
need in Appendix~\ref{app:subtractions}. Note that even after this subtraction
the threshold and large mass expansions of the two-loop form factors still receive
contributions proportional to a single logarithm $L_s\equiv \ln(-4z)$ from diagrams
with massless cuts. We therefore split the subtracted two-loop form factors into a
constant and a logarithmic part and construct separate approximants for
each part.

The top mass dependence is contained in the variable $z$ and the
conformal transformation~\cite{Fleischer:1994ef}
\begin{equation}
 z = \frac{4\omega}{(1+\omega)^2}
 \label{eq:conf_map}
\end{equation}
is used to map the entire complex $z$ plane onto the unit disc $|\omega|\leq1$
with the branch cut for $z\geq1$ corresponding to the perimeter. Thus, the top-mass
dependence is encoded by a function that is analytic in the region $|\omega|<1$
and can be reconstructed using Pad\'e approximants
\begin{equation}
  [n/m](\omega) = \frac{\sum\limits_{i=0}^n a_i \omega^i}{1 + \sum\limits_{j=1}^m b_j \omega^j}\,,
  \label{eq:Pade_ansatz}
\end{equation}
where the $n+m+1$ coefficients $a_i, b_j$ can be fixed by imposing the
condition that the expansion of eq.~\eqref{eq:Pade_ansatz} in the LME and
threshold region must reproduce the known coefficients for given, fixed
values of $r_Z$ and $\tilde{x}$. The small-mass behavior discussed in
sec.~\ref{sec:sme} is not used to further constrain the Pad\'e
coefficients, but is taken into account by a rescaling of the Pad\'e
ansatz. Hence, we use approximation functions of the form
\begin{align}
  P_{AA-VV}^{(1)}(\omega) ={}& \frac{[n/m](\omega)}{1+ a_{R,0} z(\omega)}\,,\nonumber\\
  P_{AA-VV}^{(2)}(\omega) ={}& \frac{[n/m](\omega)}{1+ a_{R,0} z(\omega)} + \frac{[k/l](\omega)}{1+ a_{R,1} z(\omega)}L_s\nonumber\\
                             & +s_{AA}^{(2)}(z(w)) - s_{VV}^{(2)}(z(w))\,,\nonumber\\
  P_{VV}^{(1)}(\omega) ={}& \frac{z(\omega) [n/m](\omega)}{1+ a_{R,0} z(\omega)}\,,\nonumber\\
  P_{VV}^{(2)}(\omega) ={}& \frac{z(\omega) [n/m](\omega)}{1+ a_{R,0} z(\omega)} + \frac{z(\omega) [k/l](\omega)}{1+ a_{R,1} z(\omega)}L_s\nonumber\\
                          & + s_{VV}^{(2)}(z(w))\,,
  \label{eq:rescaling}
\end{align}
\begin{figure*}
\begin{center}
\includegraphics[width=0.49 \textwidth]{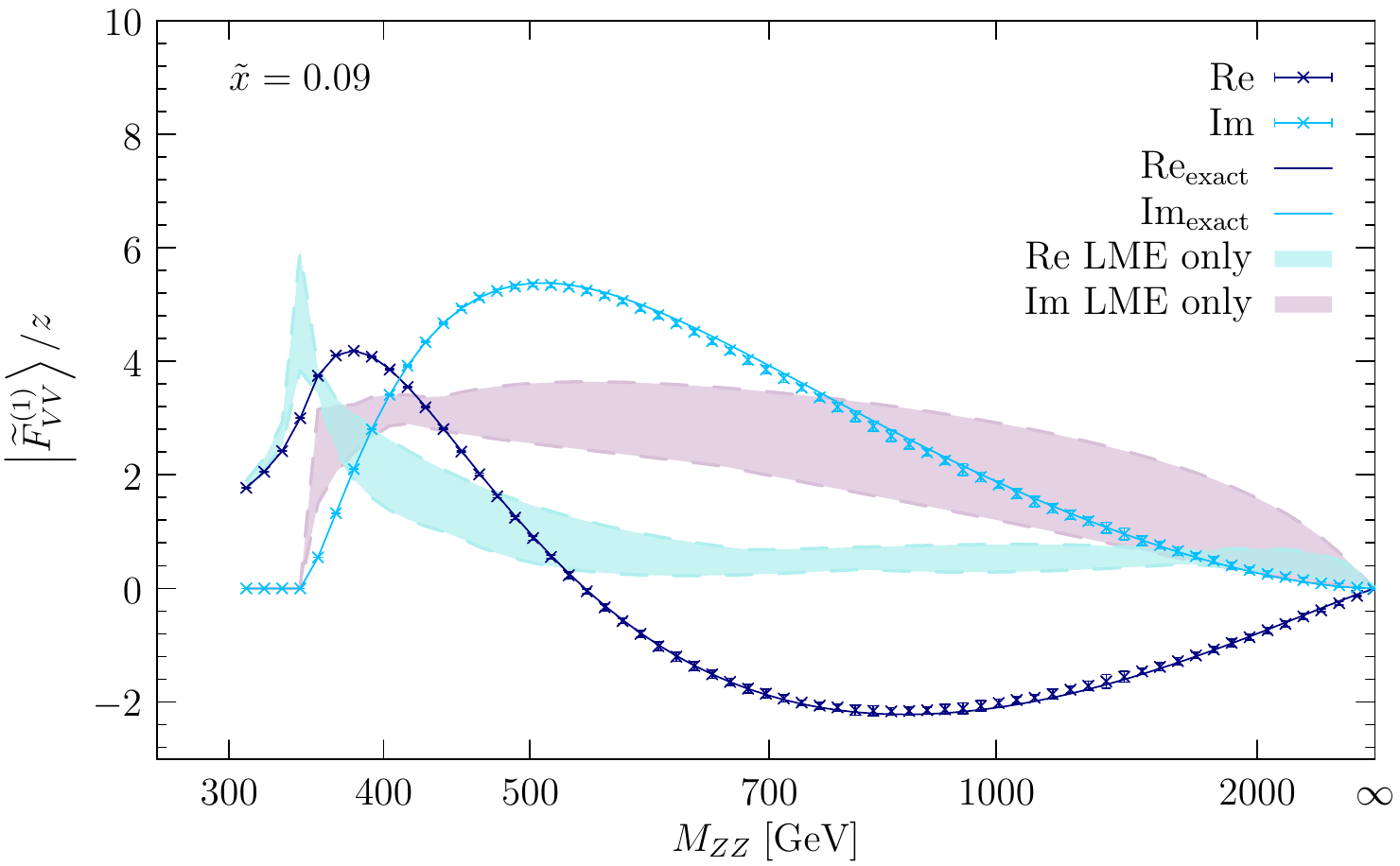}
\includegraphics[width=0.49 \textwidth]{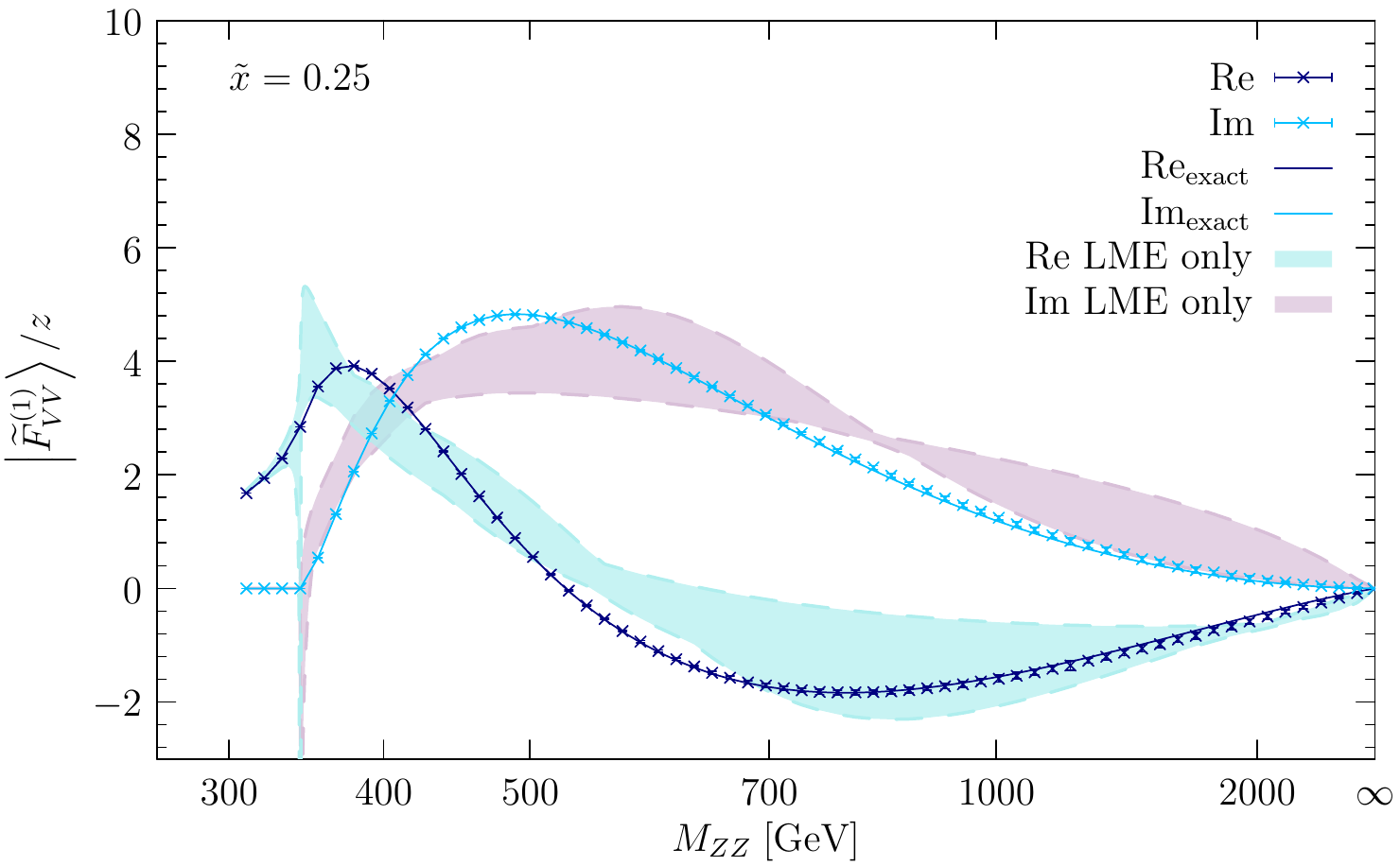}\\
\includegraphics[width=0.49 \textwidth]{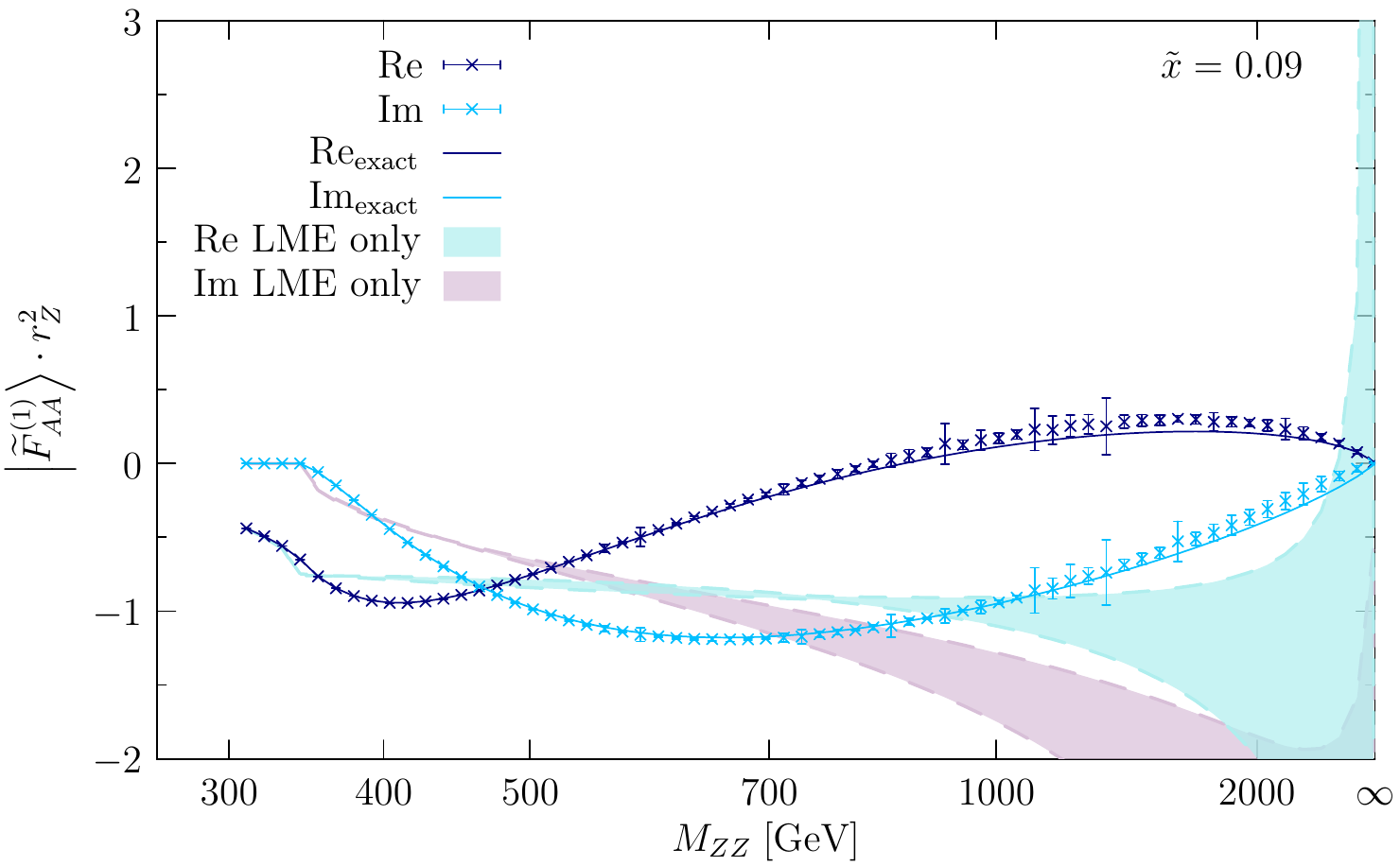}
\includegraphics[width=0.49 \textwidth]{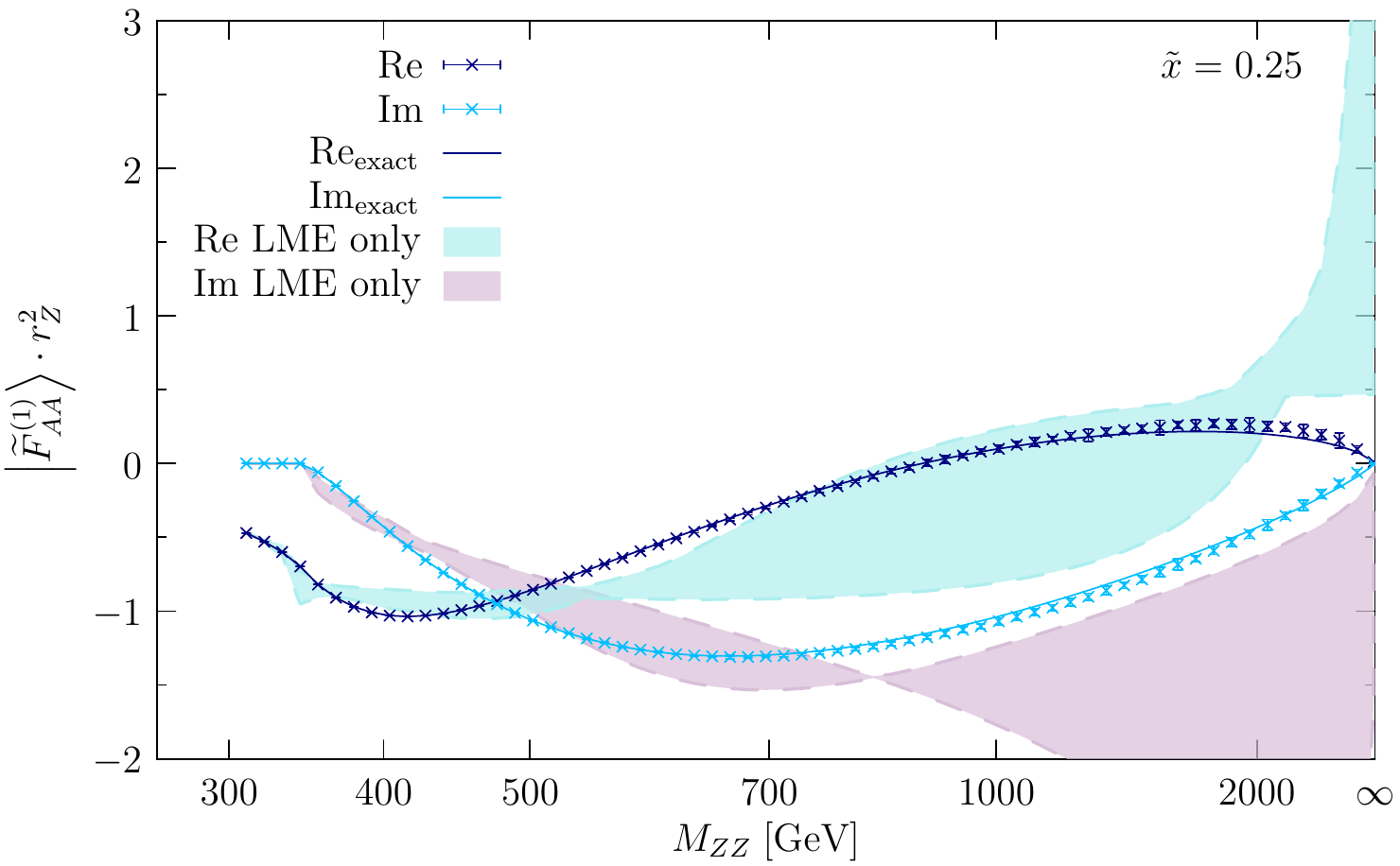}
\end{center}
\caption{ The form factors $\Ket{\widetilde{F}_{VV}^{(1)}}$ (upper row) and
$\Ket{\widetilde{F}_{AA}^{(1)}}$ (lower row) at LO for  $\tilde{x}=0.09$ (left side) and
$\tilde{x}=0.25$ (right side) as a function of the invariant mass of the
$Z$-boson pair. $\tilde{x}=0.25$ corresponds to the maximum possible
transverse momentum for a given invariant mass. The dark blue and light blue
points correspond to the real and imaginary parts of the Pad\'e approximants from
eq.~\eqref{eq:approx_AA} and eq.~\eqref{eq:approx_VV}, the solid lines are the full
result and the shaded regions are Pad\'e approximants that were constructed using
only the information from the LME (cf. text for details).\label{fig:LO}}
\end{figure*}
where $P_{AA-VV}^{(j)}$ is used to approximate the difference between
the axial-vector and vector form factors, whereas the vector form
factors in isolation are approximated using $P_{VV}^{(j)}$. The limit
$z\to\infty$ corresponds to $\omega\to-1$ where the approximants in
eq.~\eqref{eq:Pade_ansatz} approach a constant value. Thus, the rescaling
eq.~\eqref{eq:rescaling} enforces the correct asymptotic behavior for
$z\to\infty$ discussed in sec.~\ref{sec:sme} and provides us with
free parameters $a_{R,i}$ that can be varied in addition to the polynomial
degrees $n,m,k$ and $l$ to assess the stability of the approximation. We
note that these variations are performed independently for all the terms in
eq.~\eqref{eq:rescaling}. Our final ansätze for the form factor approximation
are then
\begin{align}
  \label{eq:approx_AA}
    &\Ket{\widetilde{F}_{AA}^{(j)}(z(\omega))}
    \simeq{}  P_{AA-VV}^{(j)}(\omega) + P_{VV}^{(j)}(\omega)\,,\\
  \label{eq:approx_VV}
  &\Ket{\widetilde{F}_{VV}^{(j)}(z(\omega))}
    \simeq  P_{VV}^{(j)}(\omega)\,.
\end{align}

\section{Results\label{sec:numerics}}

\begin{figure*}
\begin{center}
\includegraphics[width=0.49 \textwidth]{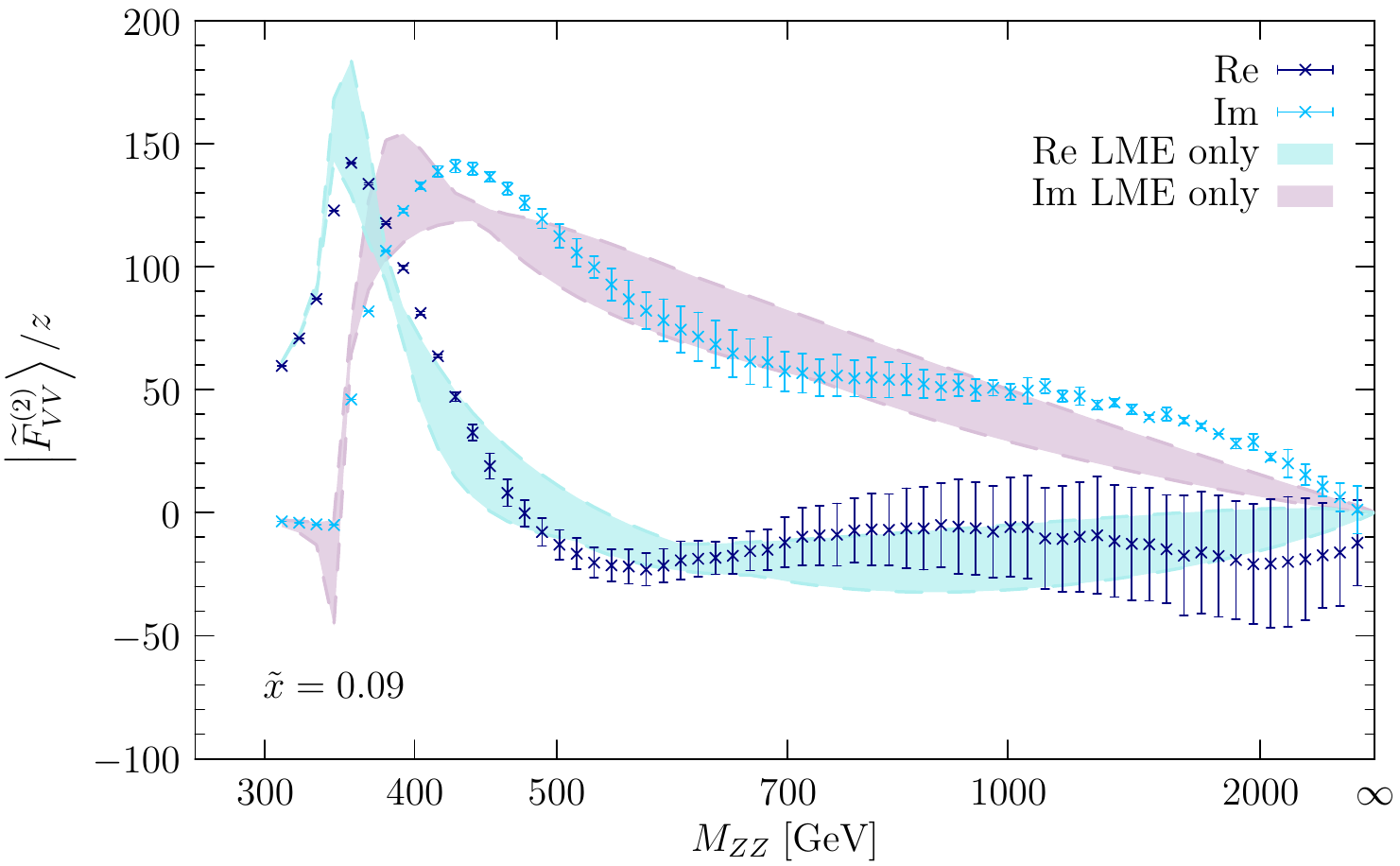}
\includegraphics[width=0.49 \textwidth]{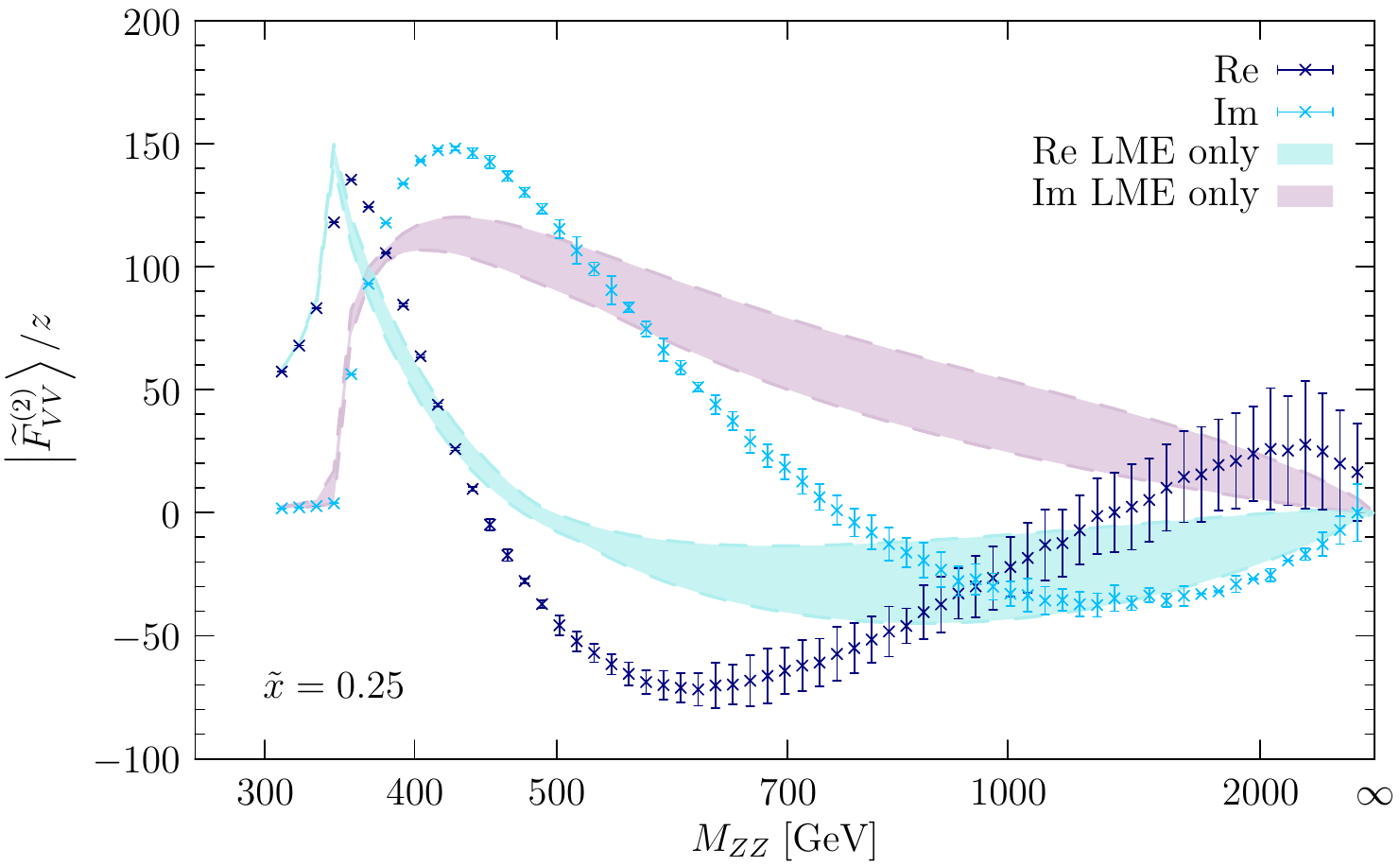}\\
\includegraphics[width=0.49 \textwidth]{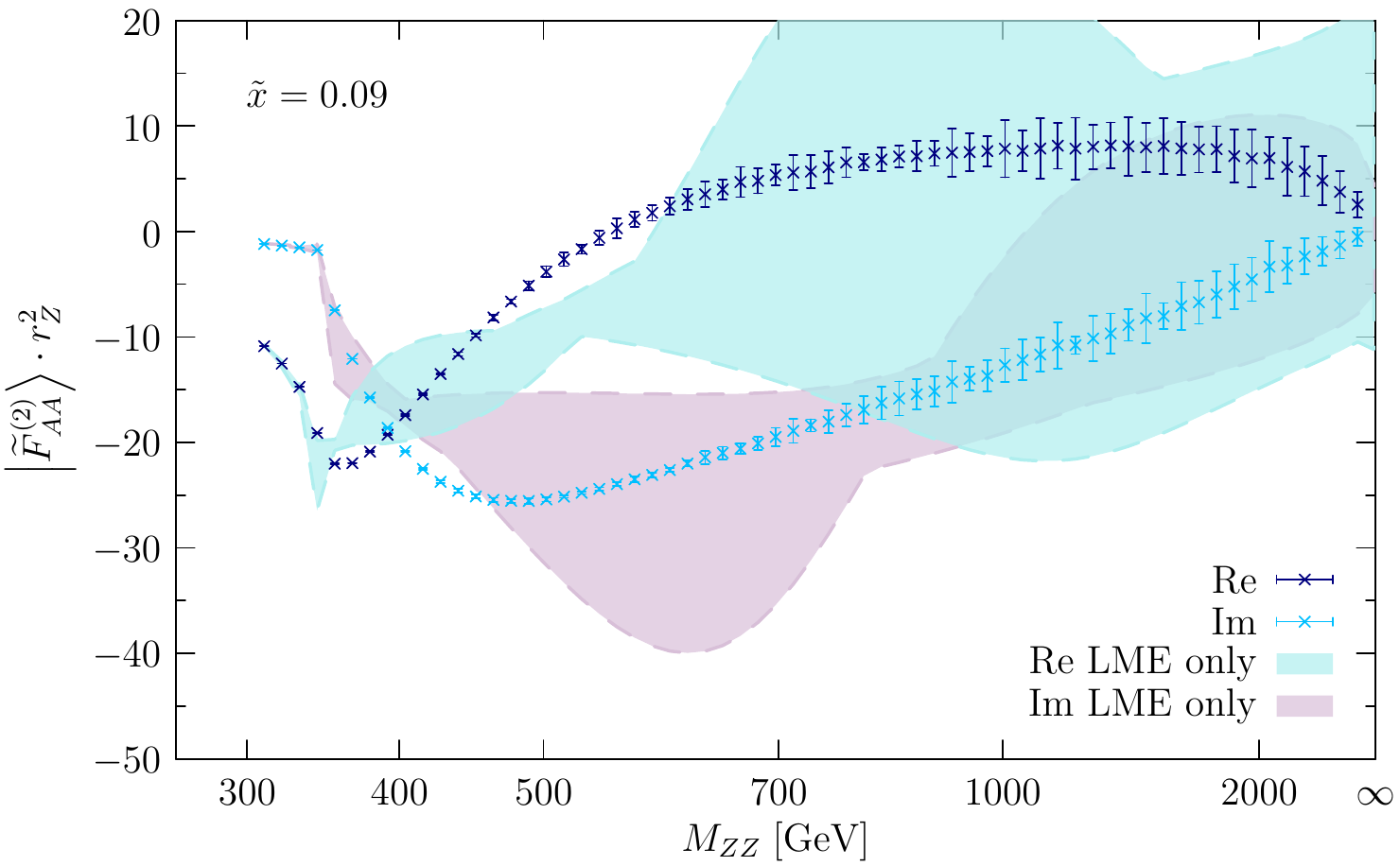}
\includegraphics[width=0.49 \textwidth]{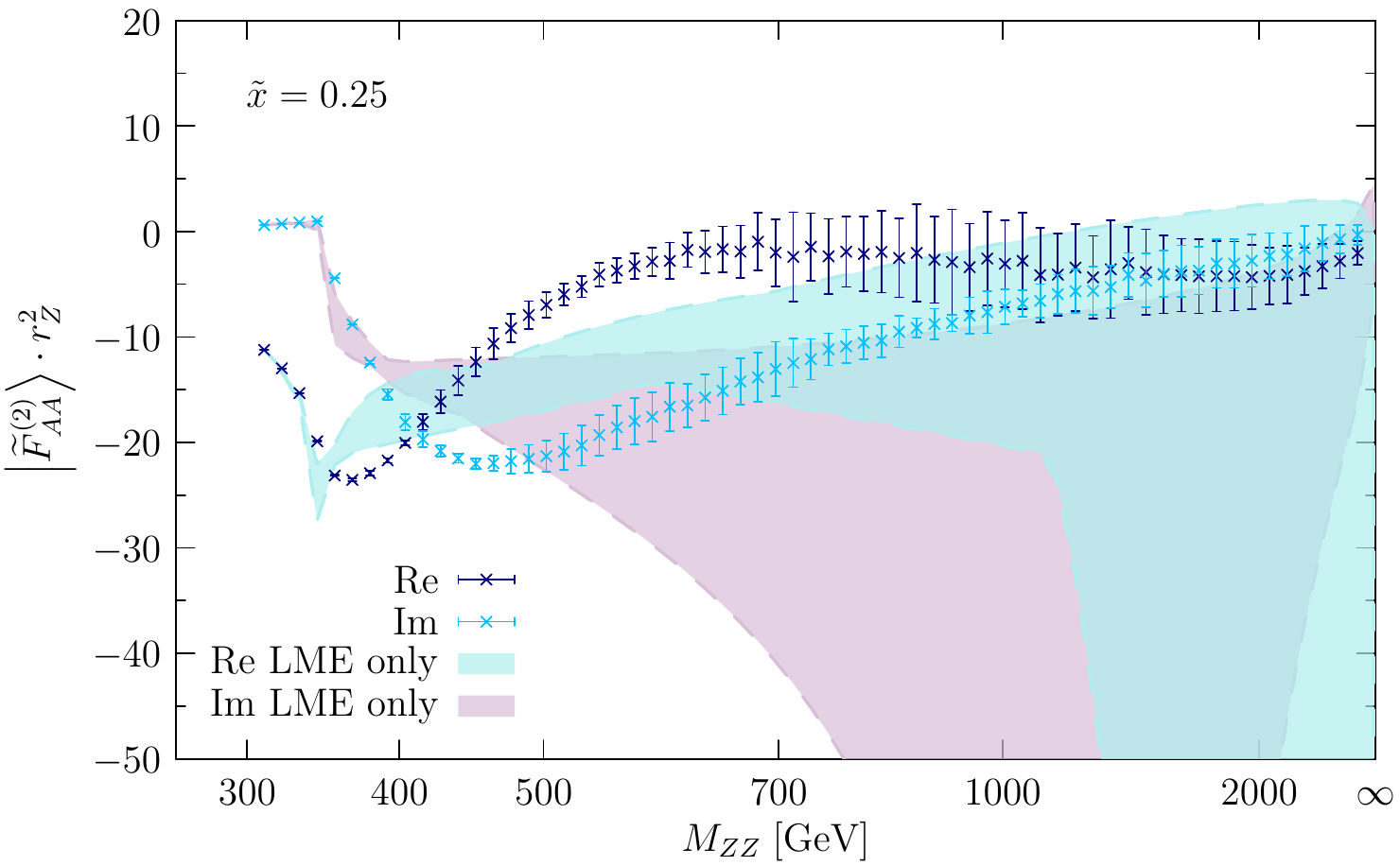}
\end{center}
\caption{ The NLO form factors $\Ket{\widetilde{F}_{VV}^{(2)}}$ (upper row) and
$\Ket{\widetilde{F}_{AA}^{(2)}}$  (lower row) for  $\tilde{x}=0.09$ (left side) and
$\tilde{x}=0.25$ (right side) as a function of the invariant mass of the
$Z$-boson pair. The conventions are the same as in fig.~\ref{fig:LO} with
the points and shaded regions corresponding to the Pad\'e approximation
constructed from the LME only.\label{fig:NLO}}
\end{figure*}

Before showing our results at NLO for the form factors, we can compare
the LO form factors constructed as discussed in the previous sections with the full
analytic result.
We choose as input for the on-shell $Z$-boson and top quark masses
\begin{equation}
m_Z=91.1876\text{ GeV}\,, \quad  m_t=173\text{ GeV}\,,
\end{equation}
and show results for two different values of $\tilde{x}$ in fig.~\ref{fig:LO}.
The plots contain the maximum information we have available from the LME
at LO (see~\cite{Campbell:2016ivq}) and our threshold expansion.
By construction the Pad\'e ansatz in eq.~\eqref{eq:Pade_ansatz} contains poles
in the complex $\omega$ plane whereas the functions it approximates are
analytic in $z$ implying the absence of poles in the unit disc $|\omega|\leq1$.
Furthermore poles in the vicinity of the unit disc can cause unphysical
behavior in the reconstructed form factors. We were not able to construct
Pad\'e approximants without poles inside a larger disc $|\omega|\leq1.2$.
Therefore we focus on the time-like region of the form factors and construct
only Pad\'e approximants which do not contain poles for
\begin{equation}
\text{Re}(z(\omega))>0 \qquad \text{and} \qquad |\omega|<1.2\,.
\label{eq:pole_criterion}
\end{equation}
We obtain an uncertainty estimate for our results in the following way.
For every phase space point, we calculate the mean and standard
deviation for each contributing Pad\'e approximant in eq.~\eqref{eq:rescaling}.
To this end, we vary the rescaling parameters $a_{R,i}$ in the region
\begin{equation}
a_{R,i} \in [0.1, 10]\,,
\end{equation}
and vary $[n/m]$ within $|n-m| \leq 3$, where $n + m + 1$ is the number
of available constraints. We construct 100 variants for each Padé
approximant. Our final prediction then follows from the sum of the mean
values of the Padé approximants, with an uncertainty obtained by adding
the individual errors in quadrature.

\begin{figure*}
\begin{center}
\includegraphics[width=0.49 \textwidth]{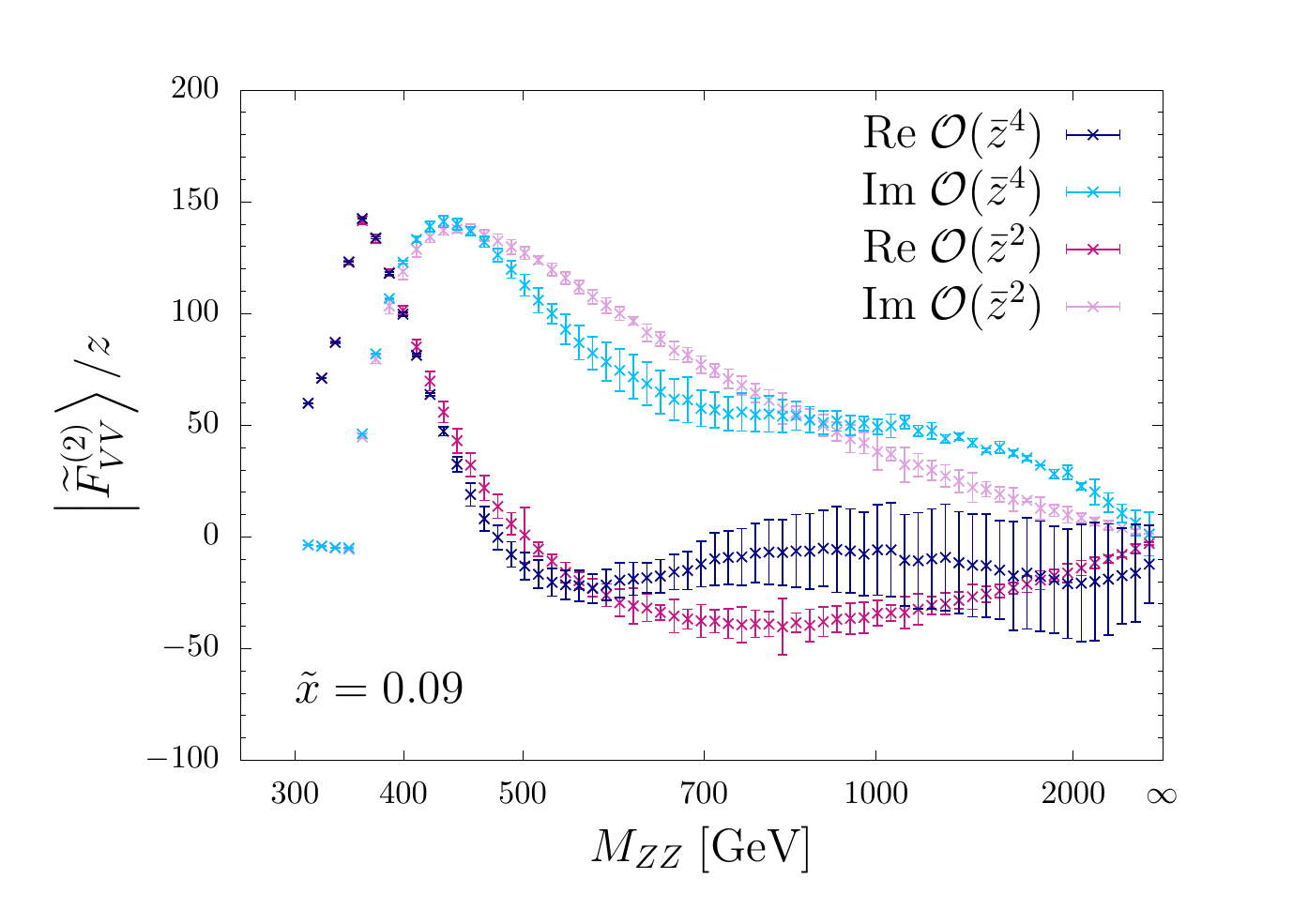}
\includegraphics[width=0.49 \textwidth]{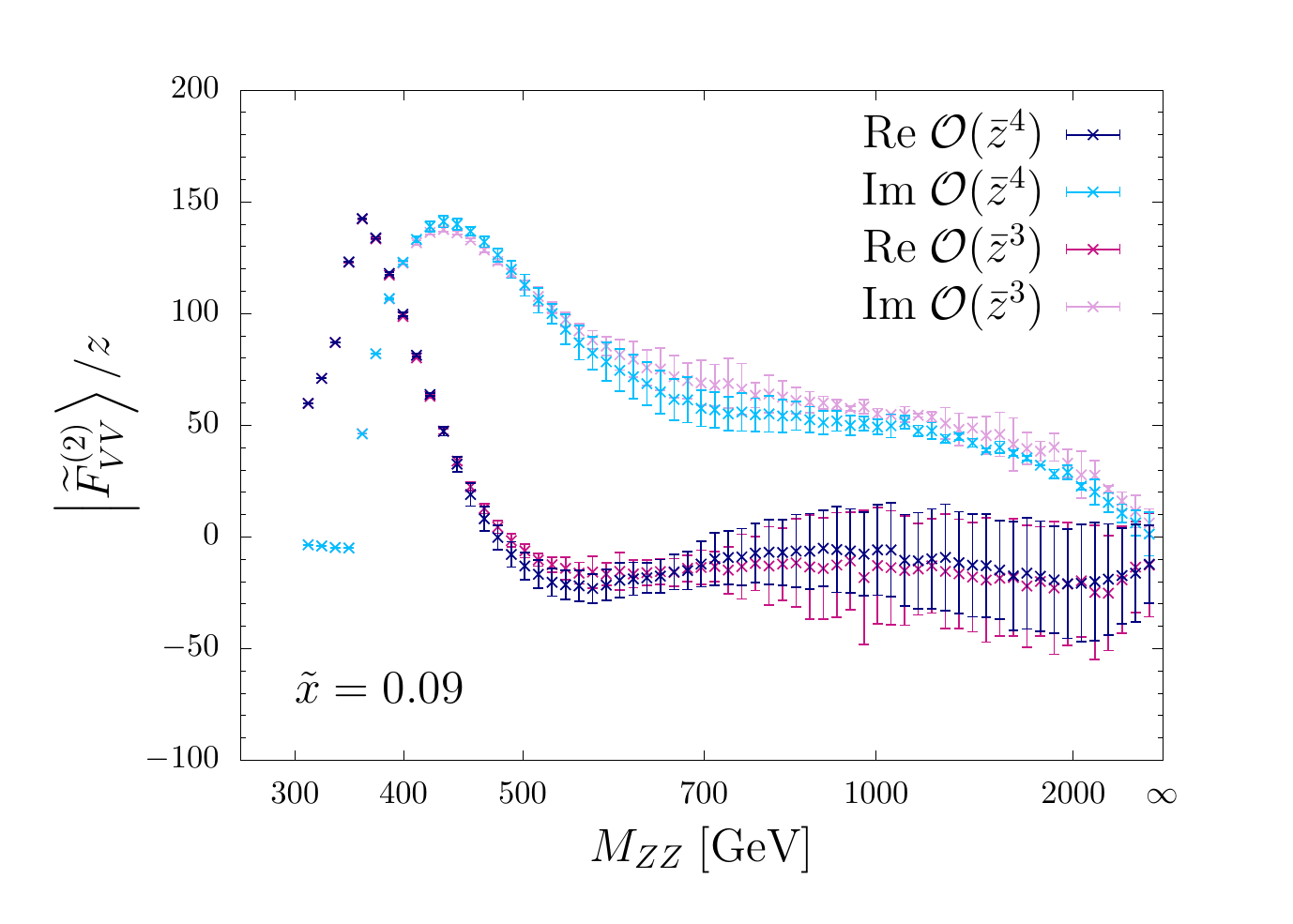}\\
\includegraphics[width=0.49 \textwidth]{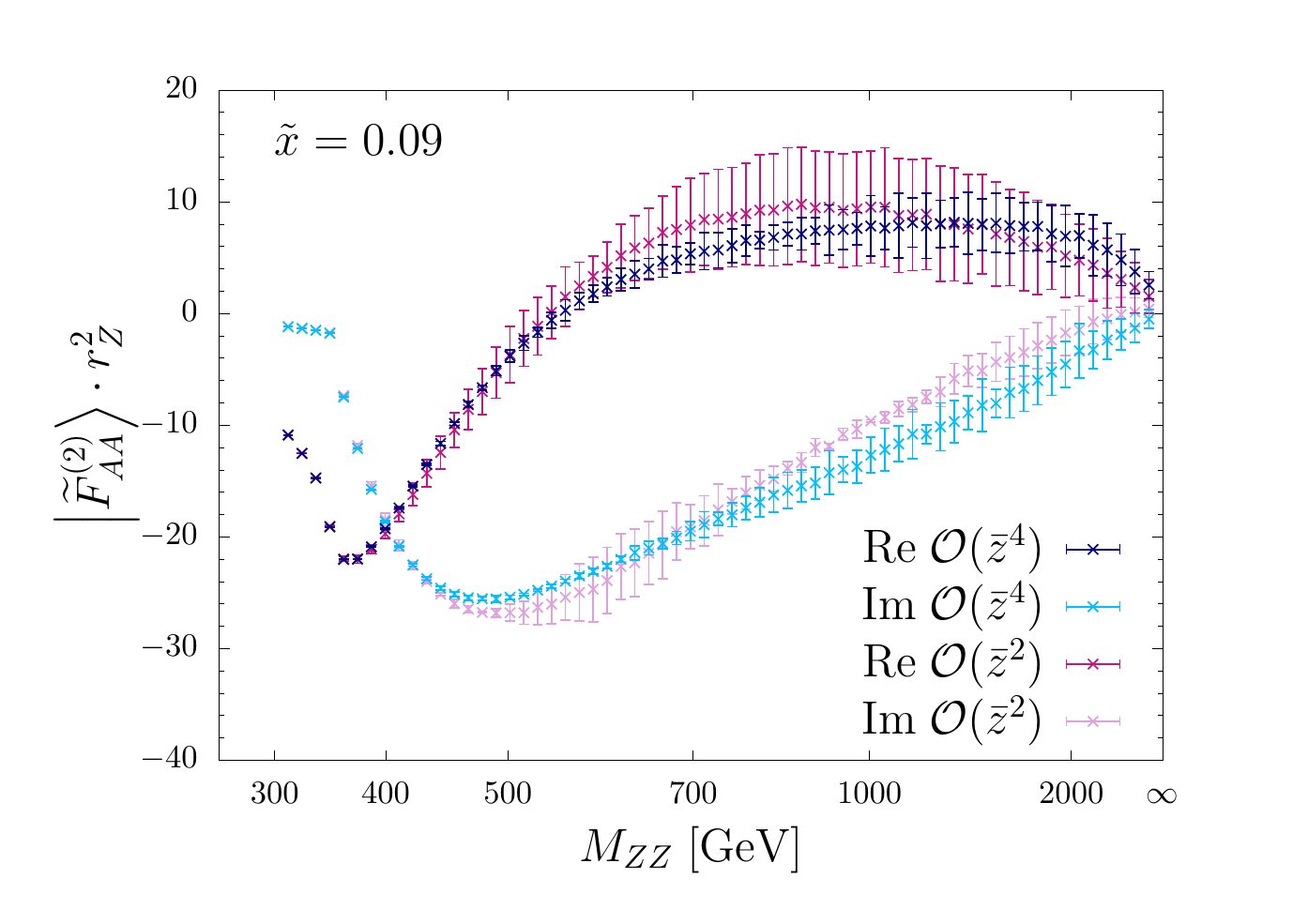}
\includegraphics[width=0.49 \textwidth]{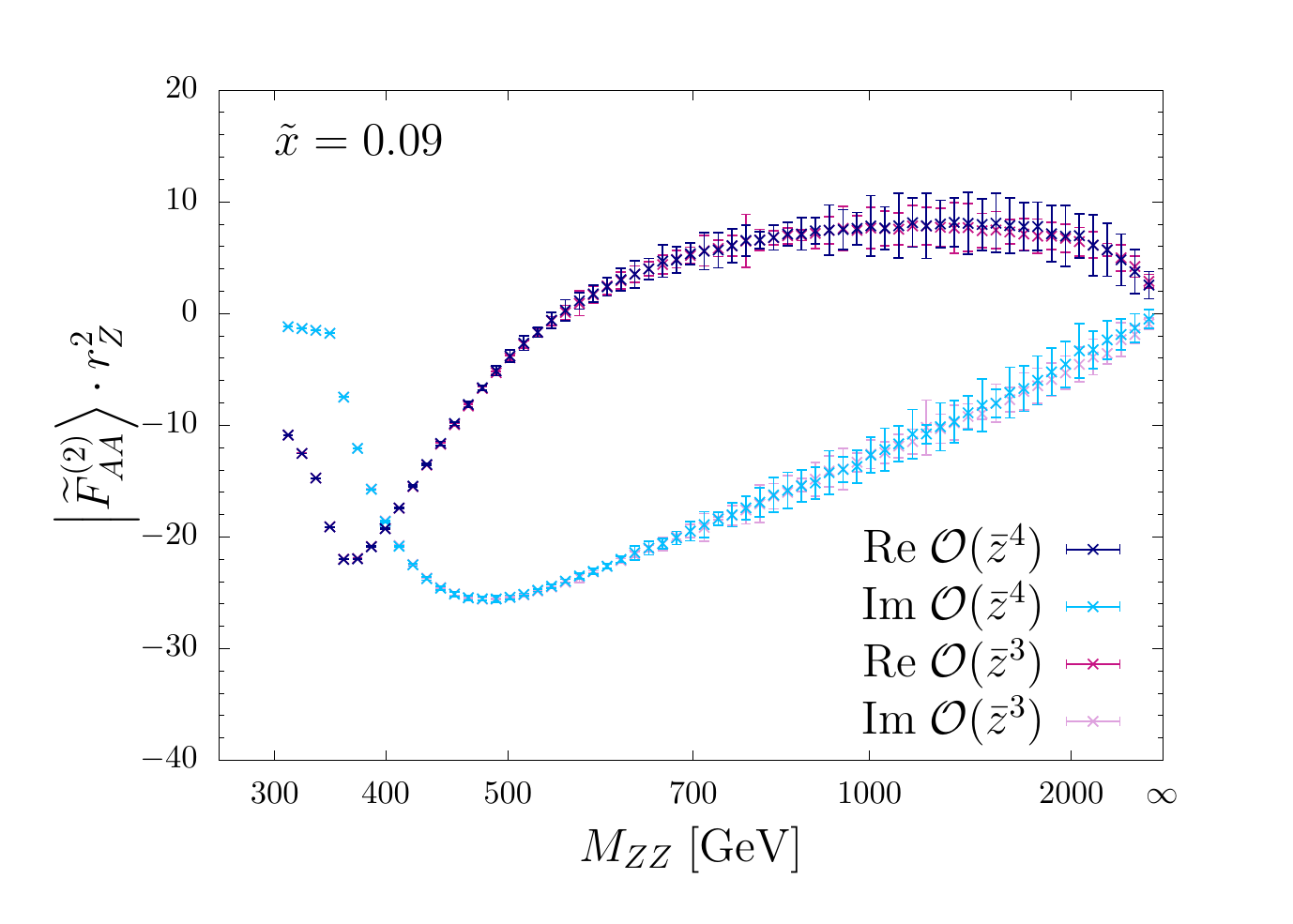}
\end{center}
\caption{ The NLO form factors $\Ket{\widetilde{F}_{VV}^{(2)}}$ (upper row) and
$\Ket{\widetilde{F}_{AA}^{(2)}}$  (lower row) for  $\tilde{x}=0.09$ as a 
function of the invariant mass of the $Z$-boson pair. In dark/light blue we show 
the same points as in fig.~\ref{fig:NLO} while in pink/rose we show the 
real/imaginary part of the Pad\'e approximants expanded up to 
$\mathcal{O}(\bar{z}^2)$ (left side) and $\mathcal{O}(\bar{z}^3)$ (right side). 
\label{fig:NLO_diff_order}}
\end{figure*}

Fig.~\ref{fig:LO} shows the Pad\'e approximants from eq.~\eqref{eq:approx_AA} and
eq.~\eqref{eq:approx_VV} for the LO form factors $\Ket{\widetilde{F}_{VV}^{(1)}}$
and $\Ket{\widetilde{F}_{AA}^{(1)}}$ including our uncertainty estimate
as points with error bars. We observe good agreement with the full results,
which are indicated by the solid lines, up to large values of the invariant mass
$M_{ZZ}$ of the $Z$-boson pair. The error remains small throughout the
whole invariant mass range, increasing somewhat towards large $M_{ZZ}$.
The behavior for different values of $\tilde{x}$ is similar. To demonstrate
the importance of including the threshold expansion we also show an
approximation based solely on the LME as shaded regions. For this we
adopt the prescription given in ref.~\cite{Campbell:2016ivq} and show
the envelope of the [2/2], [2/3], [3/2] and [3/3] Pad\'e approximants
which we have constructed without applying the rescaling of eq.~\eqref{eq:rescaling}
or the pole criterion eq.~\eqref{eq:pole_criterion}. We note that the resonant
structure near $z=1$ in the upper right plot showing the vector form factor for
maximal transverse momentum is caused by a pole near $w=1$ in the [3/3] Pad\'e
approximant. In our full results from eq.~\eqref{eq:approx_AA} and eq.~\eqref{eq:approx_VV}
we apply the criterion eq.~\eqref{eq:pole_criterion} to exclude approximants which
feature such resonances in the time-like region $z\geq0$.
We conclude that the threshold expansion is essential for the reconstruction
of the full top mass dependence above the top quark threshold.

We now turn to the NLO form factors. In fig.~\ref{fig:NLO} we show the results
for the virtual corrections to the form factors $\Ket{\widetilde{F}_{VV}^{(2)}}$
(upper panel) and $\Ket{\widetilde{F}_{AA}^{(2)}}$ (lower panel) for two values of
$\tilde{x}$. Note that we do not include the double-triangle contribution to the
form factors, as they have been computed analytically in \cite{Campbell:2016ivq}.
As at LO, we include only the top quark contributions.
The uncertainty associated with the Pad\'e construction increases with
$M_{ZZ}$. Since we input information mainly at low $M_{ZZ}$ this
behavior is expected. With the exception of the vector form factor
$\Ket{\widetilde{F}_{VV}^{(2)}}$ for small transverse momenta (upper
left panel in fig.~\ref{fig:NLO}) we find that the Pad\'e approximation
based on the LME alone does not yield a realistic reconstruction of the
top-quark mass effects of the form factors. In particular, the important
axial-vector form factor suffers from very large uncertainties. We remark though
that in \cite{Campbell:2016ivq} for the NLO cross section the Pad\'e prediction
was improved by a reweighting with the full LO cross section.

\begin{figure*}
\begin{center}
\includegraphics[width=0.49 \textwidth]{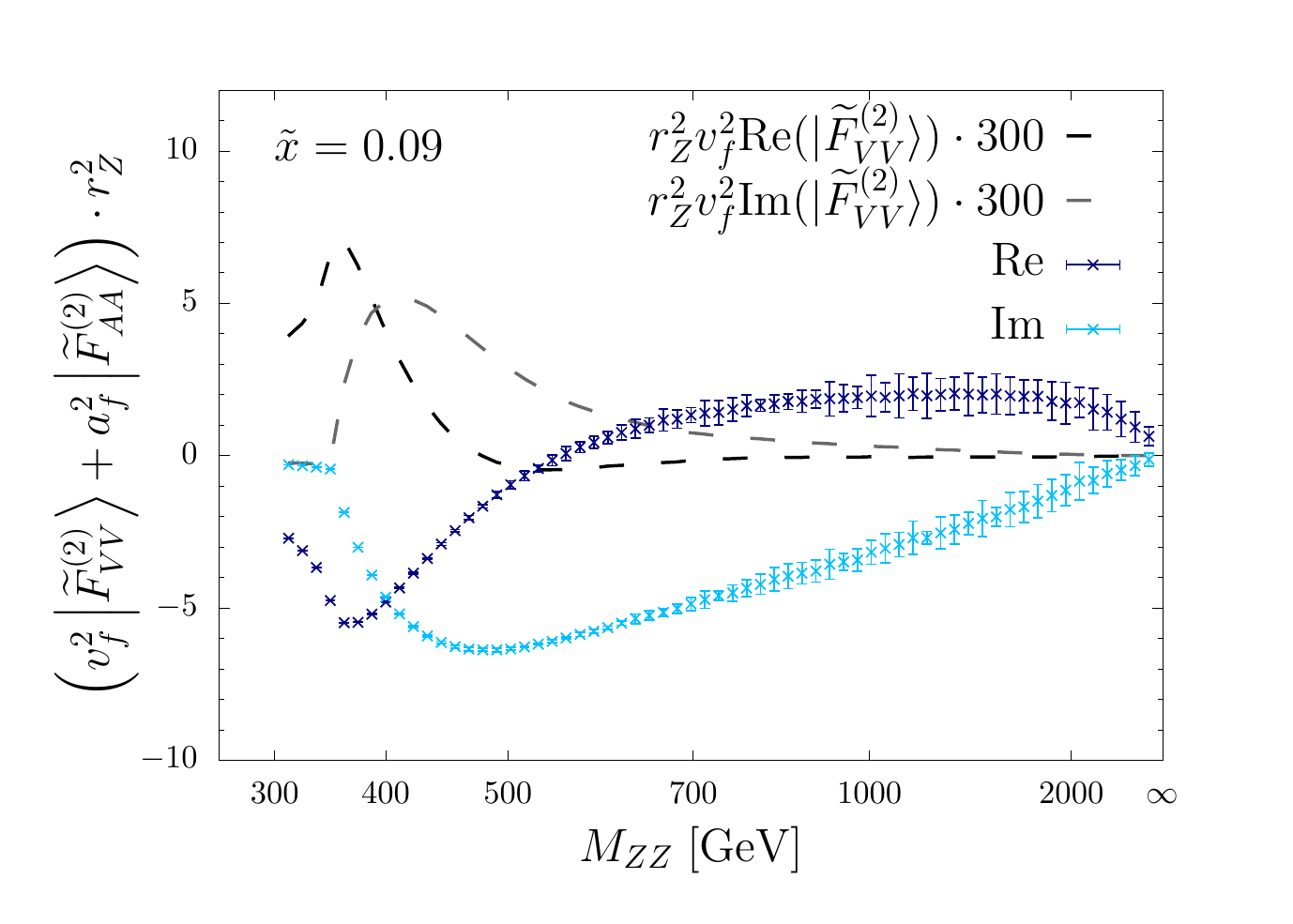}
\includegraphics[width=0.49 \textwidth]{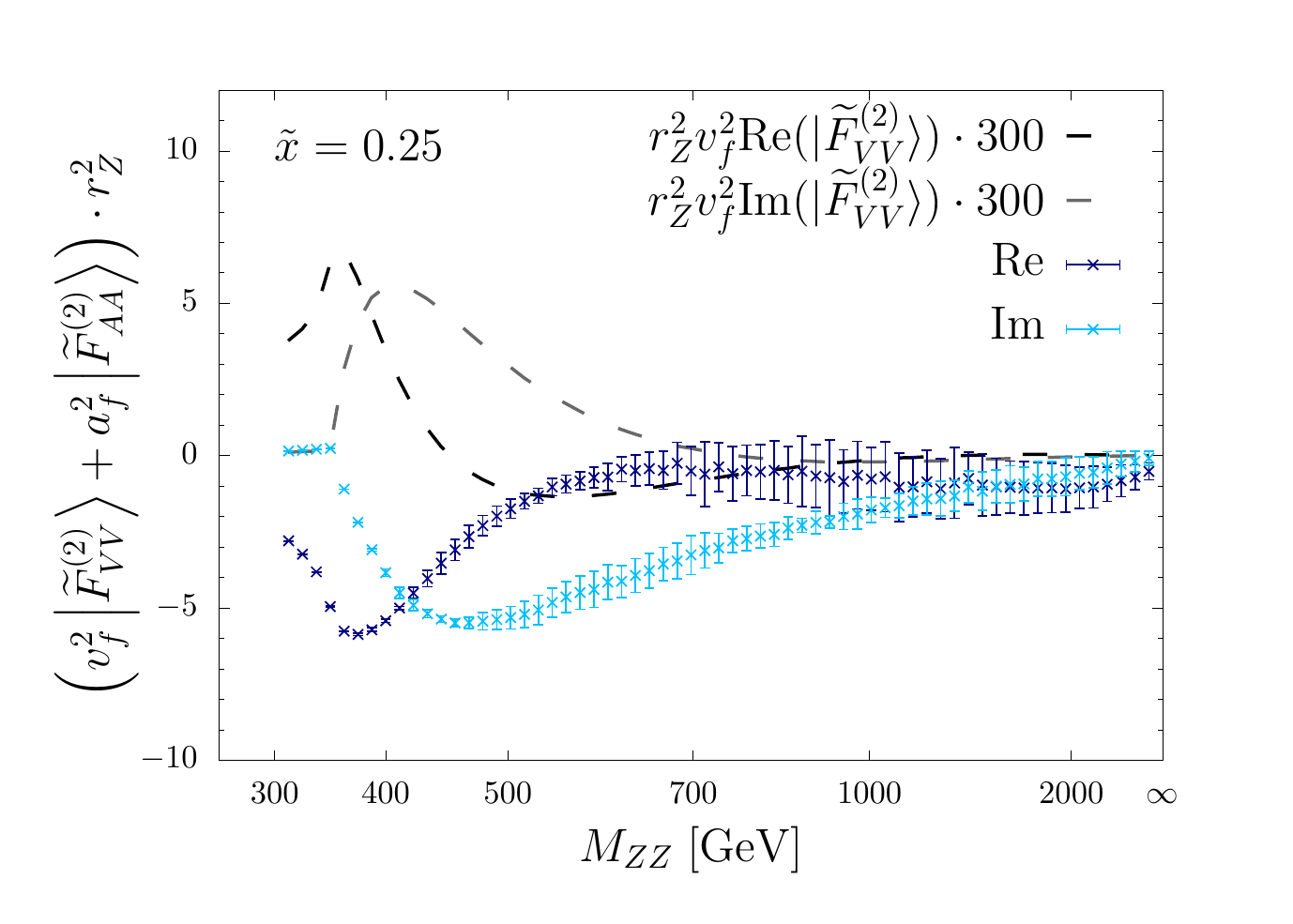}\\
\end{center}
\caption{ The interference form factor
$v_f^2\Ket{\widetilde{F}_{VV}^{(2)}}+a_f^2 \Ket{\widetilde{F}_{VV}^{(2)}} $
for  $\tilde{x}=0.09$ (left side) and $\tilde{x}=0.25$ (right side) as a
function of the invariant mass of the $Z$-boson pair. The dashed lines
show a rescaled form factor $\Ket{\widetilde{F}_{VV}^{(2)}}$ to demonstrate
that it is negligible compared to $\Ket{\widetilde{F}_{AA}^{(2)}}$. \label{fig:Vfin}}
\end{figure*}

We note that $\Ket{\widetilde{F}_{VV}^{(2)}}$ shows a small oscillation in the
region of large $M_{ZZ}$ when the transverse momentum of the $Z$ bosons is small
as is evident from the upper left plot in fig.~\ref{fig:NLO}. We trace the
appearance of the second peak back to the contribution proportional to $L_s$
stemming from diagrams with massless cuts. In general, we find that this
contribution shows worse convergence behavior than the non-logarithmic terms
when including more and more terms in the LME and the threshold expansion.
This is shown in fig.~\ref{fig:NLO_diff_order} where we compare our results
from fig.~\ref{fig:NLO} to the Pad\'e approximants obtained with the same
procedure but only using threshold input up to the order $\bar{z}^2$ and 
$\bar{z}^3$. We observe good convergence in the case of the axial-vector form 
factor. On the other hand, the $\mathcal{O}(\bar{z}^2)$ approximation for the 
vector form factor does not feature the oscillatory behavior described above 
and there is no overlap with the full approximation in a significant part of 
the phase space. 
However, the $\mathcal{O}(\bar{z}^3)$ and $\mathcal{O}(\bar{z}^4)$ results
are in good agreement with the full approximation where we have also included
the $\mathcal{O}(\bar{z}^5)$ term in the coefficient of the logarithm $L_s$ to
verify that this stabilization persists with the addition of higher orders in
the threshold expansion. We conclude from this discussion, that the Pad\'e 
approximation can be improved systematically when including higher orders 
in the various expansions. 
Nevertheless, we believe that the prediction for $\Ket{\widetilde{F}_{VV}^{(2)}}$
should be taken with a grain of salt above $M_{ZZ}\geq 500\text{ GeV}$ 
because of the slower convergence. 

In fig.~\ref{fig:Vfin} we show the virtual 
corrections to the form factor $v_f^2\Ket{\widetilde{F}_{VV}^{(2)}}+a_f^2
\Ket{\widetilde{F}_{VV}^{(2)}} $ as it enters in the interference term
with the Higgs boson exchange. The dashed lines show the form factor
$v_f^2\Ket{\widetilde{F}_{VV}^{(2)}}$ increased by a factor of 300. This
clearly demonstrates that the interference term will be dominated by
$\Ket{\widetilde{F}_{AA}^{(2)}}$ and we therefore choose not to modify
the uncertainty estimate for the vector form factor. The fact that
$\Ket{\widetilde{F}_{VV}^{(2)}}$ is negligible compared to
$\Ket{\widetilde{F}_{AA}^{(2)}}$ allows us to make trustworthy
predictions for the interference with the Higgs production with
subsequent decay to $Z$ bosons up to $M_{ZZ}\to \infty$, even though as
stated above we trust our results for $\Ket{\widetilde{F}_{VV}^{(2)}}$
only for $M_{ZZ}\le 500\text{ GeV}$.

The numerical implementation of the form factors is available as a {\tt
FORTRAN} routine on request and can be combined with existing
computations of the massless loop contributions and the real corrections
for the interference of the Higgs exchange with decay to $ZZ$ with the
continuum background.

\section{Conclusions and outlook\label{sec:conclusion}}

We have considered top-quark mass effects in the continuum process $gg
\to ZZ$, focusing on the form factors relevant for the NLO interference
with the production of a Higgs boson and its subsequent decay into two
$Z$ bosons. We have presented a Pad\'e-based approximation using
information from an expansion around a large top quark mass and an
expansion around the top quark pair production threshold.

At LO, we have shown that our Pad\'e construction approximates very well
the full top mass dependence of the form factors for the whole range of
the invariant mass $M_{ZZ}$ of the $Z$ bosons.
At NLO, we provide a new prediction with very small uncertainties at
small and moderate $M_{ZZ}$, with an increased uncertainty towards large
$M_{ZZ}$.
We expect that adding more information into the Pad\'e construction at
large $M_{ZZ}$ would improve the description also in this region.

Our results can be combined both with virtual corrections mediated by
massless loops and the real corrections. The latter constitute a
one-loop process and can therefore be computed with well-established
techniques. The Pad\'e construction can also be applied to the remaining
form factors contributing to $gg\to ZZ$, which do not interfere with the
Higgs signal.

We note also that while in this work we have applied our method to the
production of on-shell $Z$ bosons, there is no obstruction for applying
it also to off-shell $Z$ boson production. Indeed, the LME for off-shell
$Z$ boson production is already known up to the order
$z^4$~\cite{Caola:2016trd}. While a calculation of the full top mass
dependence for on-shell $Z$ bosons with numerical methods seems to be
feasible with current techniques in a reasonable time-frame
(see~\cite{talk, fullZZ}) a computation of the off-shell form factors appears to
be beyond the current state-of-the-art.

\subsubsection*{Acknowledgements}

We thank B.~Agarwal, A.~von Manteuffel and N.~Kauer for their comments on the manuscript.
This work has received funding from the European Union's Horizon 2020
research and innovation programme under the Marie Skłodowska-Curie grant
agreement No. 764850, SAGEX.
RG is supported by the ``Berliner Chancengleichheitsprogramm''.

\appendix

\section{Threshold expansion of form factors}
\label{sec:F_thr_res}

In the following we give explicit expressions for the coefficients in
the threshold expansions of the form factors. For convenience, we quote
the definition already given in eq.~\eqref{eq:formfactor_expansion}:
\begin{align}
\label{eq:formfactor_expansion_app}
 \Ket{\widetilde{\mathcal{F}}_i^{(1)}} & \mathop{\asymp}\limits^{z\to1}
    \sum_{n=3}^\infty a_i^{(n,0)}\bar{z}^{\frac{n}{2}}\,,\\
 \Ket{\widetilde{\mathcal{F}}_i^{(2)}} & \mathop{\asymp}\limits^{z\to1}
    \sum_{n=2}^\infty \,\sum_{m=\bar{n}_2}^{1} \left[b_i^{(n,m)} +
    b_{i,\ln}^{(n,m)}\ln(-4z)\right]\bar{z}^{\frac{n}{2}}\ln^m\bar{z}\,,
\end{align}
where $i \in \{VV, AA\}$ and $\bar{n}_2$ is $n$ modulo $2$. The
coefficients $a,b$ are most conveniently written in terms of the two
dimensionless ratios $r_Z = \tfrac{m^2_Z}{M^2_{ZZ}}$ and $r_{p_T} =
\tfrac{p_T^2}{M^2_{ZZ}}= \tilde{x}-r_Z$. We define the loop integral
measure as
\begin{equation}
  \label{eq:int_measure}
  [dl] = \frac{d^dl}{i\pi^{\frac{d}{2}}}e^{\epsilon \gamma_E}\,,
\end{equation}
and use the short-hand notation
\begin{equation}
  \label{eq:C_0}
  C_0 =\int [dl]\, \frac{1}{l^2[(l+q)^2-1][(l+q-p_Z)^2-1]}
\end{equation}
with $q^2=1, p_Z^2 = 4\*r_Z^2, q\cdot p_Z = 1$. The coefficients $b_{i,\ln}^{(n,1)}$ and
$b_{i,\ln}^{(2n,m)}$ vanish. Furthermore, coefficients with $m=0$ and
even $n$ do not contribute to the imaginary part and are therefore not
listed here. We have calculated the remaining coefficients $a_i^{n,0},
b_i^{n,m}$ up to $n=8$ and the coefficients $b_{i,\ln}^{(n,0)}$ up to
$n=9$, obtaining the following results:
\begin{widetext}
  \allowdisplaybreaks
  \begin{align}
    \label{eq:a_AA_3}
    a_{AA}^{(3,0)} ={}&
    \frac{4\*\pi}{3\*(1-2\*r_Z)^2\*r_Z^2}\*(-1+6\*r_Z-18\*r_Z^2+16\*r_Z^3)\,,\displaybreak[0]\\
    \label{eq:a_AA_5}
    a_{AA}^{(5,0)} ={}&\frac{2\*\pi}{15\*(1-2\*r_Z)^4\*r_Z^2}\*\big[-21+210\*r_Z-958\*r_Z^2+2336\*r_Z^3-2968\*r_Z^4+1472\*r_Z^5+8\*r_{p_T}\*(1-2\*r_Z)^2\*(1-2\*r_Z+4\*r_Z^2)\big]\,,\displaybreak[0]\\
    \label{eq:a_AA_7}
    a_{AA}^{(7,0)} ={}&\frac{\pi}{210\*(1-2\*r_Z)^6\*r_Z^2}\*\big[-905+12670\*r_Z-80954\*r_Z^2+301104\*r_Z^3-695264\*r_Z^4+985120\*r_Z^5\notag\\
    &-788896\*r_Z^6+269568\*r_Z^7+16\*r_{p_T}\*(1-2\*r_Z)^2\*(39-234\*r_Z+672\*r_Z^2-920\*r_Z^3+592\*r_Z^4)\big]\,,\displaybreak[0]\\
    \label{eq:b_AA_2_1}
    b_{AA}^{(2,1)}={}& \frac{32\*\pi^2}{9\*(1-2\*r_Z)^2\*r_Z^2}\*(-1+6\*r_Z-18\*r_Z^2+16\*r_Z^3)\,,\displaybreak[0]\\
    \label{eq:b_AA_3}
    b_{AA}^{(3,0)}
    ={}&-\frac{\pi}{9\*(1-2\*r_Z)^3\*(1-4\*r_Z)^2\*r_Z^2}\*\bigg[-2\*(1-2\*r_Z)\*(1-4\*r_Z)^2\*\big[-136+3\*\pi^2+168\*\ln(2)\big]\*r_{p_T}\*(1-2\*r_Z+4\*r_Z^2)\notag\\
    &-64\*C_0\*(1-2\*r_Z)^2\*(1-4\*r_Z)^2\*(-1-7\*r_Z+34\*r_Z^2-44\*r_Z^3+8\*r_Z^4)\notag\\
    &+64\*r_Z\*(1-4\*r_Z)^2\*\sqrt{\frac{1-r_Z}{r_Z}}\*\arctan\bigg(\frac{2\*\sqrt{(1-r_Z)\*r_Z}}{1-2\*r_Z}\bigg)\*(9-45\*r_Z+70\*r_Z^2-56\*r_Z^3+32\*r_Z^4)\notag\\
    &-(1-2\*r_Z)\*(1-4\*r_Z)\*\big[192+9\*\pi^2-56\*\ln(2)-1728\*r_Z-90\*\pi^2\*r_Z+560\*\ln(2)\*r_Z+6256\*r_Z^2+384\*\pi^2\*r_Z^2\notag\\
    &\qquad-2016\*\ln(2)\*r_Z^2-12480\*r_Z^3-816\*\pi^2\*r_Z^3+3584\*\ln(2)\*r_Z^3+12032\*r_Z^4+576\*\pi^2\*r_Z^4-3584\*\ln(2)\*r_Z^4-2048\*r_Z^5\big]\notag\\
    &+128\*(1-2\*r_Z)\*\ln(2-4\*r_Z)\*(2-23\*r_Z+98\*r_Z^2-184\*r_Z^3+152\*r_Z^4-112\*r_Z^5+96\*r_Z^6)\bigg]
      \,,\displaybreak[0]\\
    \label{eq:b_AA_3_0_ln}
b_{AA,\ln}^{(3,0)}
    ={}&0\,,\displaybreak[0]\\
    \label{eq:b_AA_3_1}
b_{AA}^{(3,1)}
    ={}&0\,,\displaybreak[0]\\
    \label{eq:b_AA_4_1}
    b_{AA}^{(4,1)}={}& \frac{32\*\pi^2}{45\*(1-2\*r_Z)^4\*r_Z^2}\*\big[-3+30\*r_Z-134\*r_Z^2+328\*r_Z^3-464\*r_Z^4+256\*r_Z^5+4\*(1-2\*r_Z)^2\*r_{p_T}\*(1-2\*r_Z+4\*r_Z^2)\big]\,,\displaybreak[0]\\
    \label{eq:b_AA_5}
    b_{AA}^{(5,0)}
    ={}&\frac{\pi}{4050\*(1-2\*r_Z)^5\*(1-4\*r_Z)^3\*r_Z^2}\*\bigg[
         -(1-2\*r_Z)\*(1-4\*r_Z)\*\big[21472-6075\*\pi^2-104520\*\ln(2)-643776\*r_Z\notag\\
    &\qquad+109350\*\pi^2\*r_Z+1881360\*\ln(2)\*r_Z+7528432\*r_Z^2-855900\*\pi^2\*r_Z^2-14848320\*\ln(2)\*r_Z^2-44282176\*r_Z^3\notag\\
    &\qquad+3817800\*\pi^2\*r_Z^3+67172160\*\ln(2)\*r_Z^3+141881152\*r_Z^4-10558080\*\pi^2\*r_Z^4-187708800\*\ln(2)\*r_Z^4\notag\\
    &\qquad-245787136\*r_Z^5+18135360\*\pi^2\*r_Z^5+319941120\*\ln(2)\*r_Z^5+202149888\*r_Z^6-17763840\*\pi^2\*r_Z^6\notag\\
    &\qquad-302008320\*\ln(2)\*r_Z^6-35688448\*r_Z^7+7326720\*\pi^2\*r_Z^7+117350400\*\ln(2)\*r_Z^7-25067520\*r_Z^8\big]\notag\\
    &-18\*(1-2\*r_Z)\*(1-4\*r_Z)\*r_{p_T}\*\big[2136+375\*\pi^2-360\*\ln(2)-37584\*r_Z-5250\*\pi^2\*r_Z+5040\*\ln(2)\*r_Z\notag\\
    &\qquad+263040\*r_Z^2+30000\*\pi^2\*r_Z^2-28800\*\ln(2)\*r_Z^2-915648\*r_Z^3-94200\*\pi^2\*r_Z^3+22080\*\ln(2)\*r_Z^3\notag\\
    &\qquad+1546624\*r_Z^4+184560\*\pi^2\*r_Z^4+424320\*\ln(2)\*r_Z^4-1103872\*r_Z^5-218880\*\pi^2\*r_Z^5\notag\\
    &\qquad-1320960\*\ln(2)\*r_Z^5+546816\*r_Z^6+111360\*\pi^2\*r_Z^6+768000\*\ln(2)\*r_Z^6-163840\*r_Z^7\big]\notag\\
    &+C_0\*(1-2\*r_Z)\*(1-4\*r_Z)\*\big[46080\*(1-2\*r_Z)^3\*r_{p_T}\*(2-9\*r_Z+20\*r_Z^2-12\*r_Z^3-8\*r_Z^4+16\*r_Z^5)\notag\\
    &\qquad-2880\*(1-2\*r_Z)\*(1-4\*r_Z)\*(31-99\*r_Z-1242\*r_Z^2+8912\*r_Z^3-23696\*r_Z^4+29840\*r_Z^5-16608\*r_Z^6+2176\*r_Z^7)\big]\notag\\
    &+r_Z\*(1-4\*r_Z)\*\sqrt{\frac{1-r_Z}{r_Z}}\*\arctan\bigg(\frac{2\*\sqrt{(1-r_Z)\*r_Z}}{1-2\*r_Z}\bigg)\*\big[960\*(1-4\*r_Z)\*(-687+8843\*r_Z\notag\\
    &\qquad\qquad-46162\*r_Z^2+126356\*r_Z^3-195416\*r_Z^4+174304\*r_Z^5-95360\*r_Z^6+30720\*r_Z^7)\notag\\
    &\qquad-46080\*(1-2\*r_Z)^2\*r_{p_T}\*(-6+45\*r_Z-152\*r_Z^2+264\*r_Z^3-176\*r_Z^4+16\*r_Z^5)\big]\notag\\
    &+(1-2\*r_Z)\*\ln(2-4\*r_Z)\*\big[1920\*(-170+3251\*r_Z-26282\*r_Z^2+117196\*r_Z^3-314896\*r_Z^4\notag\\
    &\qquad\qquad+524464\*r_Z^5-549440\*r_Z^6+394688\*r_Z^7-240512\*r_Z^8+96768\*r_Z^9)\notag\\
    &\qquad-92160\*(1-2\*r_Z)^2\*r_{p_T}\*(-1+13\*r_Z-70\*r_Z^2+200\*r_Z^3-280\*r_Z^4+128\*r_Z^5)\big]\bigg]
      \,,\displaybreak[0]\\
    \label{eq:b_AA_5_0_ln}
b_{AA,\ln}^{(5,0)}
    ={}&\frac{16\*\pi}{5\*(1-2\*r_Z)^2\*r_Z^2}\*(-1+4\*r_Z+6\*r_{p_T})\*(1-2\*r_Z+4\*r_Z^2)\,,\displaybreak[0]\\
    \label{eq:b_AA_5_1}
b_{AA}^{(5,1)}
    ={}&\frac{32\pi}{135\*(1-2\*r_Z)^2\*r_Z^2}\*\big[53-318\*r_Z+846\*r_Z^2-848\*r_Z^3-108\*r_{p_T}\*(1-2\*r_Z+4\*r_Z^2)\big]\,,\displaybreak[0]\\
    \label{eq:b_AA_6_1}
b_{AA}^{(6,1)}
    ={}&\frac{32\*\pi^2}{315\*(1-2\*r_Z)^6\*r_Z^2}\*\big[19-266\*r_Z+1730\*r_Z^2-6504\*r_Z^3+14648\*r_Z^4-19072\*r_Z^5+12128\*r_Z^6-2816\*r_Z^7\notag\\
    &+4\*(1-2\*r_Z)^2\*r_{p_T}\*(9-54\*r_Z+168\*r_Z^2-208\*r_Z^3+128\*r_Z^4)\big]\,,\displaybreak[0]\\
    \label{eq:b_AA_7}
    b_{AA}^{(7,0)}
    ={}&\frac{\pi}{793800\*(1-r_Z)\*(1-2\*r_Z)^7\*(1-4\*r_Z)^4\*r_Z^2}\*\bigg[(1-r_Z)\*(1-2\*r_Z)\*(1-4\*r_Z)\*\big[-48296976+1306935\*\pi^2\notag\\
    &\qquad+68546520\*\ln(2)+1381163616\*r_Z-33980310\*\pi^2\*r_Z-1782209520\*\ln(2)\*r_Z-19953987184\*r_Z^2\notag\\
    &\qquad+396287640\*\pi^2\*r_Z^2+20934631200\*\ln(2)\*r_Z^2+178890721728\*r_Z^3-2743009920\*\pi^2\*r_Z^3\notag\\
    &\qquad-146825159040\*\ln(2)\*r_Z^3-1038970811008\*r_Z^4+12546379440\*\pi^2\*r_Z^4+682278602880\*\ln(2)\*r_Z^4\notag\\
    &\qquad+3951328802304\*r_Z^5-39832823520\*\pi^2\*r_Z^5-2194951852800\*\ln(2)\*r_Z^5-9798614287104\*r_Z^6\notag\\
    &\qquad+89300171520\*\pi^2\*r_Z^6+4941762577920\*\ln(2)\*r_Z^6+15459576751104\*r_Z^7-140100468480\*\pi^2\*r_Z^7\notag\\
    &\qquad-7665966120960\*\ln(2)\*r_Z^7-14595082354688\*r_Z^8+147641840640\*\pi^2\*r_Z^8+7817008496640\*\ln(2)\*r_Z^8\notag\\
    &\qquad+7124726562816\*r_Z^9-94379765760\*\pi^2\*r_Z^9-4715940741120\*\ln(2)\*r_Z^9-1110704586752\*r_Z^{10}\notag\\
    &\qquad+27358248960\*\pi^2\*r_Z^{10}+1270490726400\*\ln(2)\*r_Z^{10}-100576788480\*r_Z^{11}\big]\notag\\
    &+2\*(1-r_Z)\*(1-2\*r_Z)\*(1-4\*r_Z)\*r_{p_T}\*\big[12875128-2482515\*\pi^2-32397960\*\ln(2)-322712656\*r_Z\notag\\
    &\qquad+54615330\*\pi^2\*r_Z+712755120\*\ln(2)\*r_Z+3481217888\*r_Z^2-530889660\*\pi^2\*r_Z^2-7045775520\*\ln(2)\*r_Z^2\notag\\
    &\qquad-21890762240\*r_Z^3+3016742400\*\pi^2\*r_Z^3+41557608960\*\ln(2)\*r_Z^3+89967740544\*r_Z^4-11128456080\*\pi^2\*r_Z^4\notag\\
    &\qquad-162222883200\*\ln(2)\*r_Z^4-251512872192\*r_Z^5+27788412960\*\pi^2\*r_Z^5+434414211840\*\ln(2)\*r_Z^5\notag\\
    &\qquad+477015472640\*r_Z^6-47130431040\*\pi^2\*r_Z^6-795771594240\*\ln(2)\*r_Z^6-592554727424\*r_Z^7\notag\\
    &\qquad+52573812480\*\pi^2\*r_Z^7+963913574400\*\ln(2)\*r_Z^7+437616467968\*r_Z^8-35230325760\*\pi^2\*r_Z^8\notag\\
    &\qquad-705638277120\*\ln(2)\*r_Z^8-137703424000\*r_Z^9+10818662400\*\pi^2\*r_Z^9+235343216640\*\ln(2)\*r_Z^9\notag\\
    &\qquad-8257536000\*r_Z^{10}\big]\notag\\
    &+201600\*(1-2\*r_Z)^3\*(1-4\*r_Z)^4\*\big[-356+3\*\pi^2+520\*\ln(2)\big]\*r_{p_T}^2\*(-1+r_Z)\*(1-2\*r_Z+4\*r_Z^2)\notag\\
    &+C_0\*(1-r_Z)\*(1-2\*r_Z)^2\*(1-4\*r_Z)^2\*\big[-20160\*(1-4\*r_Z)\*(1773-13705\*r_Z-15494\*r_Z^2+542860\*r_Z^3\notag\\
    &\qquad\qquad-2672760\*r_Z^4+6652528\*r_Z^5-9490464\*r_Z^6+7595840\*r_Z^7-2952064\*r_Z^8+311808\*r_Z^9)\notag\\
    &\qquad+645120\*(1-2\*r_Z)^2\*r_{p_T}\*(96-601\*r_Z+1800\*r_Z^2-4344\*r_Z^3+7592\*r_Z^4-5664\*r_Z^5-32\*r_Z^6+1600\*r_Z^7)\big]\notag\\
    &+r_Z\*(1-4\*r_Z)^2\*\sqrt{\frac{1-r_Z}{r_Z}}\*\arctan\bigg(\frac{2\*\sqrt{(1-r_Z)\*r_Z}}{1-2\*r_Z}\bigg)\*\big[-6720\*(1-4\*r_Z)\*(35337-616496\*r_Z\notag\\
    &\qquad\qquad+4722401\*r_Z^2-20896358\*r_Z^3+59069448\*r_Z^4-111353552\*r_Z^5+142045200\*r_Z^6\notag\\
    &\qquad\qquad-122427488\*r_Z^7+70734336\*r_Z^8-26438656\*r_Z^9+5122048\*r_Z^{10})\notag\\
    &\qquad-215040\*(1-r_Z)\*(1-2\*r_Z)^2\*r_{p_T}\*(-1044+10253\*r_Z-43216\*r_Z^2\notag\\
    &\qquad\qquad+104204\*r_Z^3-158912\*r_Z^4+150832\*r_Z^5-70976\*r_Z^6+9408\*r_Z^7)\big]\notag\\
    &-\ln(2-4\*r_Z)\*(1-r_Z)\*(1-2\*r_Z)\*\big[13440\*(9662-255657\*r_Z+3011806\*r_Z^2-20841376\*r_Z^3\notag\\
    &\qquad\qquad+94139672\*r_Z^4-291779424\*r_Z^5+635459136\*r_Z^6-982191360\*r_Z^7+1084817792\*r_Z^8\notag\\
    &\qquad\qquad-874003712\*r_Z^9+538395136\*r_Z^{10}-253452288\*r_Z^{11}+66985984\*r_Z^{12})\notag\\
    &\qquad+430080\*(1-2\*r_Z)^2\*r_{p_T}\*(-251+4211\*r_Z-30950\*r_Z^2+132676\*r_Z^3-370528\*r_Z^4\notag\\
    &\qquad\qquad+699712\*r_Z^5-871904\*r_Z^6+647168\*r_Z^7-220160\*r_Z^8+18432\*r_Z^9)
      \big]\bigg]\,,\displaybreak[0]\\
    \label{eq:b_AA_7_0_ln}
b_{AA,\ln}^{(7,0)}
    ={}&\frac{8\*\pi}{35\*(1-2\*r_Z)^4\*r_Z^2}\*(-1+4\*r_Z+6\*r_{p_T})\*(39-234\*r_Z+672\*r_Z^2-920\*r_Z^3+592\*r_Z^4)\,,\displaybreak[0]\\
    \label{eq:b_AA_7_1}
b_{AA}^{(7,1)}
    ={}&\frac{16\pi}{945\*(1-2\*r_Z)^4\*r_Z^2}\*\big[2079-20790\*r_Z+91822\*r_Z^2-218096\*r_Z^3+269368\*r_Z^4-137024\*r_Z^5\notag\\
    &-4\*r_{p_T}\*(1169-7014\*r_Z+20000\*r_Z^2-27624\*r_Z^3+17840\*r_Z^4)\big]\,,\displaybreak[0]\\
    \label{eq:b_AA_8_1}
b_{AA}^{(8,1)}
    ={}&\frac{32\*\pi^2}{945\*(1-2\*r_Z)^8\*r_Z^2}\*\big[233-4194\*r_Z+34914\*r_Z^2-174856\*r_Z^3+575264\*r_Z^4\notag\\
    &-1278272\*r_Z^5+1903168\*r_Z^6-1816064\*r_Z^7 +992768\*r_Z^8-235520\*r_Z^9\notag\\
    &+4\*(1-2\*r_Z)^2\*r_{p_T}\*(11-110\*r_Z+572\*r_Z^2-1544\*r_Z^3+2288\*r_Z^4-1184\*r_Z^5+192\*r_Z^6)\notag\\
    &-64\*(1-2\*r_Z)^4\*r_{p_T}^2\*(1-2\*r_Z+4\*r_Z^2)\big]\,,\displaybreak[0]\\
    \label{eq:b_AA_9_0_ln}
    b_{AA,\ln}^{(9,0)}
    ={}&\frac{-2\*\pi}{315\*(1-2\*r_Z)^6\*r_Z^2}\*\big[2621-36694\*r_Z+225124\*r_Z^2-793216\*r_Z^3+1727344\*r_Z^4-2317408\*r_Z^5+1757888\*r_Z^6\notag\\
    &-610560\*r_Z^7-2\*r_{p_T}\*(8183-81830\*r_Z+373652\*r_Z^2-964400\*r_Z^3+1472912\*r_Z^4-1224416\*r_Z^5+457920\*r_Z^6)\notag\\
    &+3200\*r_{p_T}^2\*(1-2\*r_Z)^2\*(1-2\*r_Z+4\*r_Z^2)\big]\,.
  \end{align}
for the expansion of the axial-vector component. The corresponding
coefficients in the expansion of the vector part read
\begin{align}
  \label{eq:a_VV_3}
  a_{VV}^{(3,0)}={}&\frac{16\*\pi}{3\*(1-2\*r_Z)^2}\*(2-5\*r_Z)\,,\displaybreak[0]\\
  \label{eq:a_VV_5}
  a_{VV}^{(5,0)}={}&\frac{8\*\pi}{15\*(1-2\*r_Z)^4}\*\big[34-217\*r_Z+492\*r_Z^2-412\*r_Z^3+8\*(1-2\*r_Z)^2\*r_{p_T}\big]\,,\displaybreak[0]\\
  \label{eq:a_VV_7}
  a_{VV}^{(7,0)}={}&\frac{2\*\pi}{105\*(1-2\*r_Z)^6}\*\big[1314-13549\*r_Z+57240\*r_Z^2-124296\*r_Z^3+141056\*r_Z^4-70032\*r_Z^5\notag\\
    &+16\*(1-2\*r_Z)^2\*r_{p_T}\*(11-72\*r_Z+148\*r_Z^2)\big]\,,\displaybreak[0]\\
  \label{eq:b_VV_2_1}
  b_{VV}^{(2,1)}={}&\frac{128\*\pi^2}{9\*(1-2\*r_Z)^2}\*(2-5\*r_Z)\,,\displaybreak[0]\\
  \label{eq:b_VV_3}
  b_{VV}^{(3,0)}={}&-\frac{\pi}{9\*(1-2\*r_Z)^3\*(1-4\*r_Z)^2}\*\bigg[4\*(1-2\*r_Z)\*(1-4\*r_Z)\*\big(136+21\*\pi^2+56\*\ln(2)-680\*r_Z\notag\\
  &\qquad-138\*\pi^2\*r_Z-448\*\ln(2)\*r_Z+480\*r_Z^2+216\*\pi^2\*r_Z^2+896\*\ln(2)\*r_Z^2+512\*r_Z^3\big)\notag\\
    &-8\*(1-2\*r_Z)\*(1-4\*r_Z)^2\*\big(-136+3\*\pi^2+168\*\ln(2)\big)\*r_{p_T}\notag\\
    &-128\*C_0\*(1-2\*r_Z)^2\*(1-4\*r_Z)^2\*(-3+4\*r_Z^2)\notag\\
    &+64\*(1-4\*r_Z)^2\*\sqrt{\frac{1-r_Z}{r_Z}}\*\arctan\bigg(\frac{2\*\sqrt{(1-r_Z)\*r_Z}}{1-2\*r_Z}\bigg)\*(-5+46\*r_Z-96\*r_Z^2+32\*r_Z^3)\notag\\
    &+512\*(1-2\*r_Z)\*\ln(2-4\*r_Z)\*(2-23\*r_Z+86\*r_Z^2-112\*r_Z^3+24\*r_Z^4)\bigg]\,,\displaybreak[0]\\
    \label{eq:b_VV_3_0_ln}
b_{VV,\ln}^{(3,0)}
    ={}&0\,,\displaybreak[0]\\
    \label{eq:b_VV_3_1}
b_{VV}^{(3,1)}
    ={}&0\,,\displaybreak[0]\\
  \label{eq:b_VV_4_1}
  b_{VV}^{(4,1)}={}
&\frac{128\*\pi^2}{45\*(1-2\*r_Z)^4}\*(2-11\*r_Z+36\*r_Z^2-56\*r_Z^3+4\*(1-2\*r_Z)^2\*r_{p_T})\,,\displaybreak[0]\\
  \label{eq:b_VV_5}
  b_{VV}^{(5,0)}={}
&-\frac{\pi}{2025\*(1-r_Z)\*(1-2\*r_Z)^5\*(1-4\*r_Z)^3}\*\bigg[-2\*(1-2\*r_Z)\*(1-4\*r_Z)\*(1-r_Z)\*\big(186008-6345\*\pi^2-177000\*\ln(2)\notag\\
&\qquad-2712968\*r_Z+88830\*\pi^2\*r_Z+2535360\*\ln(2)\*r_Z+15732800\*r_Z^2-498420\*\pi^2\*r_Z^2-14241120\*\ln(2)\*r_Z^2\notag\\
&\qquad-45725792\*r_Z^3+1443960\*\pi^2\*r_Z^3+39511680\*\ln(2)\*r_Z^3+68432384\*r_Z^4-2246400\*\pi^2\*r_Z^4\notag\\
&\qquad-55011840\*\ln(2)\*r_Z^4-45982208\*r_Z^5+1537920\*\pi^2\*r_Z^5+31488000\*\ln(2)\*r_Z^5+6266880\*r_Z^6\big)\notag\\
    &+36\*(1-2\*r_Z)\*(1-4\*r_Z)\*(1-r_Z)\*r_{p_T}\*\big(-8024+585\*\pi^2+11400\*\ln(2)+97248\*r_Z-6900\*\pi^2\*r_Z\notag\\
&\qquad-130080\*\ln(2)\*r_Z-384928\*r_Z^2+28860\*\pi^2\*r_Z^2+505440\*\ln(2)\*r_Z^2+532224\*r_Z^3-49440\*\pi^2\*r_Z^3\notag\\
&\qquad-718080\*\ln(2)\*r_Z^3-88576\*r_Z^4+27840\*\pi^2\*r_Z^4+192000\*\ln(2)\*r_Z^4-40960\*r_Z^5\big)\notag\\
&+C_0\*(1-r_Z)\*(1-2\*r_Z)^2\*(1-4\*r_Z)\*\big[2880\*(1-4\*r_Z)\*(53-564\*r_Z+2000\*r_Z^2-2368\*r_Z^3-336\*r_Z^4+1088\*r_Z^5)\notag\\
  &\qquad-46080\*(1-2\*r_Z)^2\*r_{p_T}\*(3-24\*r_Z+40\*r_Z^2+8\*r_Z^3)\big]\notag\\
  &+(1-r_Z)\*(1-2\*r_Z)\*\ln(2-4\*r_Z)\*\big[-3840\*(-80+1549\*r_Z-12538\*r_Z^2\notag\\
  &\qquad\qquad+54340\*r_Z^3-134752\*r_Z^4+187264\*r_Z^5-126752\*r_Z^6+24192\*r_Z^7)\notag\\
  &\qquad-92160\*(1-2\*r_Z)^2\*r_{p_T}\*(-1+16\*r_Z-64\*r_Z^2+80\*r_Z^3)\big]\notag\\
  &+\sqrt{\frac{1-r_Z}{r_Z}}\*\arctan\bigg(\frac{2\*\sqrt{(1-r_Z)\*r_Z}}{1-2\*r_Z}\bigg)\*\big[480\*(1-4\*r_Z)^2\*(-318+4749\*r_Z-29882\*r_Z^2\notag\\
  &\qquad\qquad+103460\*r_Z^3-212040\*r_Z^4+249152\*r_Z^5-144896\*r_Z^6+30720\*r_Z^7)\notag\\
  &\qquad+23040\*(1-2\*r_Z)^2\*(1-4\*r_Z)\*(1-r_Z)\*r_{p_T}\*(-5+54\*r_Z-184\*r_Z^2+200\*r_Z^3+16\*r_Z^4)\big]\bigg]\,,\displaybreak[0]\\
  \label{eq:b_VV_5_0_ln}
b_{VV,\ln}^{(5,0)}={}&\frac{64\*\pi}{5\*(1-2\*r_Z)^2}\*(-1+4\*r_Z+6\*r_{p_T})\,,\displaybreak[0]\\
  \label{eq:b_VV_5_1}
b_{VV}^{(5,1)}={}&\frac{128\*\pi}{135\*(1-2\*r_Z)^2}\*(-52+103\*r_Z-108\*r_{p_T})\,,\displaybreak[0]\\
  \label{eq:b_VV_6_1}
b_{VV}^{(6,1)}={}&\frac{128\*\pi^2}{315\*(1-2\*r_Z)^6}\*\big[-58+619\*r_Z-2624\*r_Z^2+5540\*r_Z^3-5552\*r_Z^4+1648\*r_Z^5+4\*(1-2\*r_Z)^2\*r_{p_T}\*(-5+6\*r_Z+32\*r_Z^2)\big]\,,\displaybreak[0]\\
  \label{eq:b_VV_7}
b_{VV}^{(7,0)}={}
  &-\frac{\pi}{198450\*(1-2\*r_Z)^7\*(1-4\*r_Z)^4\*(1-r_Z)^2}\*\bigg[-(1-2\*r_Z)\*(1-4\*r_Z)\*(1-r_Z)\*\big[122096632-127575\*\pi^2\notag\\
  &\qquad-92998920\*\ln(2)-2848934592\*r_Z+1630125\*\pi^2\*r_Z+2155938120\*\ln(2)\*r_Z+29029779592\*r_Z^2\notag\\
  &\qquad-2152710\*\pi^2\*r_Z^2-21749266560\*\ln(2)\*r_Z^2-169738450784\*r_Z^3-74541600\*\pi^2\*r_Z^3\notag\\
  &\qquad+125261747520\*\ln(2)\*r_Z^3+628426159680\*r_Z^4+590919840\*\pi^2\*r_Z^4-453823735680\*\ln(2)\*r_Z^4\notag\\
  &\qquad-1530688965120\*r_Z^5-2089568880\*\pi^2\*r_Z^5+1072845164160\*\ln(2)\*r_Z^5+2460756033152\*r_Z^6\notag\\
  &\qquad+3889861920\*\pi^2\*r_Z^6-1656277002240\*\ln(2)\*r_Z^6-2533430536704\*r_Z^7-3445787520\*\pi^2\*r_Z^7\notag\\
  &\qquad+1613260615680\*\ln(2)\*r_Z^7+1538421684224\*r_Z^8+694310400\*\pi^2\*r_Z^8-902101401600\*\ln(2)\*r_Z^8\notag\\
  &\qquad-445393100800\*r_Z^9+435456000\*\pi^2\*r_Z^9+220520939520\*\ln(2)\*r_Z^9+25144197120\*r_Z^{10}\big]\notag\\
  &-2\*(1-2\*r_Z)\*(1-4\*r_Z)\*(1-r_Z)^2\*r_{p_T}\*\big[8486296-785295\*\pi^2-11199720\*\ln(2)-222273632\*r_Z\notag\\
  &\qquad+17777340\*\pi^2\*r_Z+267660960\*\ln(2)\*r_Z+2469324096\*r_Z^2-173313000\*\pi^2\*r_Z^2-2810969280\*\ln(2)\*r_Z^2\notag\\
  &\qquad-14926447104\*r_Z^3+941371200\*\pi^2\*r_Z^3+16619420160\*\ln(2)\*r_Z^3+52801418112\*r_Z^4-3062631600\*\pi^2\*r_Z^4\notag\\
  &\qquad-58795390080\*\ln(2)\*r_Z^4-108303542784\*r_Z^5+5922262080\*\pi^2\*r_Z^5+122121377280\*\ln(2)\*r_Z^5\notag\\
  &\qquad+116805871616\*r_Z^6-6230165760\*\pi^2\*r_Z^6-134713743360\*\ln(2)\*r_Z^6-48174653440\*r_Z^7\notag\\
  &\qquad+2704665600\*\pi^2\*r_Z^7+58835804160\*\ln(2)\*r_Z^7-2064384000\*r_Z^8\big]\notag\\
  &+201600\*(1-2\*r_Z)^3\*(1-4\*r_Z)^4\*(1-r_Z)^2\*\big[-356+3\*\pi^2+520\*\ln(2)\big]\*r_{p_T}^2\notag\\
  &+C_0\*(1-r_Z)^2\*(1-2\*r_Z)^2\*(1-4\*r_Z)^2\*\big[10080\*(1-4\*r_Z)\*(1879-29708\*r_Z+186948\*r_Z^2\notag\\
  &\qquad\qquad-592848\*r_Z^3+968208\*r_Z^4-663104\*r_Z^5-57920\*r_Z^6+155904\*r_Z^7)\notag\\
  &\qquad-322560\*(1-2\*r_Z)^2\*r_{p_T}\*(85-896\*r_Z+3708\*r_Z^2-7176\*r_Z^3+5312\*r_Z^4+800\*r_Z^5)\big]\notag\\
  &+(1-4\*r_Z)^2\*\sqrt{\frac{1-r_Z}{r_Z}}\*\arctan\bigg(\frac{2\*\sqrt{(1-r_Z)\*r_Z}}{1-2\*r_Z}\bigg)\*\big[-1680\*(1-4\*r_Z)\*(14088-262050\*r_Z\notag\\
  &\qquad\qquad+2162807\*r_Z^2-10516934\*r_Z^3+33617664\*r_Z^4-74234384\*r_Z^5+114844848\*r_Z^6\notag\\
  &\qquad\qquad-121868128\*r_Z^7+82948992\*r_Z^8-31822336\*r_Z^9+5122048\*r_Z^{10})\notag\\
  &\qquad-53760\*(1-r_Z)\*(1-2\*r_Z)^2\*r_{p_T}\*(504-7341\*r_Z+44374\*r_Z^2-142180\*r_Z^3\notag\\
  &\qquad\qquad+253056\*r_Z^4-229360\*r_Z^5+68704\*r_Z^6+9408\*r_Z^7)\big]\notag\\
  &+\ln(2-4\*r_Z)\*(1-r_Z)^2\*(1-2\*r_Z)\*\big[13440\*(2522-68893\*r_Z+832786\*r_Z^2-5869624\*r_Z^3+26678552\*r_Z^4\notag\\
  &\qquad\qquad-81564240\*r_Z^5+169456096\*r_Z^6-235064576\*r_Z^7+205653632\*r_Z^8-98444288\*r_Z^9+16746496\*r_Z^{10})\notag\\
  &\qquad+215040\*(1-2\*r_Z)^2\*r_{p_T}\*(17-772\*r_Z+9460\*r_Z^2-51664\*r_Z^3\notag\\
  &\qquad\qquad+146272\*r_Z^4-212992\*r_Z^5+124160\*r_Z^6+9216\*r_Z^7)\big]\bigg]\,,\displaybreak[0]\\
  \label{eq:b_VV_7_0_ln}
b_{VV,\ln}^{(7,0)}={}
  &\frac{32\*\pi}{35\*(1-2\*r_Z)^4}\*(-1+4\*r_Z+6\*r_{p_T})\*(11-72\*r_Z+148\*r_Z^2)\,,\displaybreak[0]\\
  \label{eq:b_VV_7_1}
b_{VV}^{(7,1)}={}
&\frac{64\*\pi}{945\*(1-2\*r_Z)^4}\*\big[-2128+12817\*r_Z-25428\*r_Z^2+16060\*r_Z^3+4\*r_{p_T}\*(-413+2408\*r_Z-4460\*r_Z^2)\big]\,,\displaybreak[0]\\
  \label{eq:b_VV_8_1}
b_{VV}^{(8,1)}={}
&\frac{128\*\pi^2}{945\*(1-2\*r_Z)^8}\*\big[-430+6217\*r_Z-38796\*r_Z^2+135616\*r_Z^3-286976\*r_Z^4+367712\*r_Z^5-261632\*r_Z^6+74752\*r_Z^7\notag\\
  &+4\*r_{p_T}\*(1-2\*r_Z)^2\*(-43+290\*r_Z-764\*r_Z^2+808\*r_Z^3+48\*r_Z^4)-64\*r_{p_T}^2\*(1-2\*r_Z)^4\big]\,,\displaybreak[0]\\
  \label{eq:b_VV_9_0_ln}
b_{VV,\ln}^{(9,0)}={}
  &\frac{8\*\pi}{315\*(1-2\*r_Z)^6}\*\big[187+564\*r_Z-19112\*r_Z^2+92640\*r_Z^3-186896\*r_Z^4+152640\*r_Z^5\notag\\
  &+r_{p_T}\*(-482-12992\*r_Z+95984\*r_Z^2-233344\*r_Z^3+228960\*r_Z^4)-3200\*(1-2\*r_Z)^2\*r_{p_T}^2\big]\,.
\end{align}

\section{Subtractions \label{app:subtractions}}

In this appendix, we give the functions $s_i$ with $i \in \{VV,AA\}$
used to subtract the threshold logarithms. We write them in terms of
auxiliary subtraction functions $s_n, n \in \mathbb{N}$, i.e.
\begin{equation}
  \label{eq:s_i}
  s_i^{(2)}(z) = \sum_{n=2}^\infty C_{i,n} s_n(z)\,,
\end{equation}
where the coefficients $C_{i,n}$ are constants and $s_n \mathop{\asymp}\limits^{z\to1}
\bar{z}^{\frac{n}{2}}\ln(\bar{z}) + \mathcal{O}(\bar{z}^{\frac{n+1}{2}})$
in the threshold region. We construct these auxiliary functions based on
the known analytical results for the vacuum polarization function. The
subtraction functions and their threshold expansions are
\begin{eqnarray}
 s_2(z) & = & -\frac{16 (1-z) \Pi^{(1),v}(z)}{3 z}   \nonumber    \\
 & \mathop{\asymp}\limits^{z\to1} & (1-z)\ln(1-z) -\frac{8}{\pi} (1-z)^{3/2} + \frac13 (1-z)^2\ln(1-z)  -\frac{8}{9 \pi}\left( -5 +18 \ln(2) \right) (1-z)^{5/2}  \nonumber \\ &- & \frac{16}{3 \pi} (1-z)^{5/2}
\ln(1-z)-\frac23 (1-z)^3 \ln(1-z)  +  \frac{1}{675 \pi}\left(14653 -26280 \ln(2) \right) (1-z)^{7/2} \nonumber \\ & - &\frac{548}{45 \pi}(1-z)^{7/2} \ln(1-z)-2 (1-z)^4 \ln(1-z)+ \mathcal{O}\left((1-z)^{9/2}\right)  \label{eq:subs2}\,,\\
 s_4(z) & = & -\frac{8 }{81 \pi ^2}\,\frac{54 \pi ^2 (1-z)^2 \Pi^{(1),v}(z)-41 z}{z^2} \nonumber \\  &  \nonumber   \,\,\mathop{\asymp}\limits^{z\to1} & (1-z)^2\ln(1-z) -\frac{8}{\pi} (1-z)^{5/2} +\frac{4}{3} (1-z)^3 \ln (1-z) -\frac{16}{9 \pi}\left( 2 +\ln(512)\right)(1-z)^{7/2} \\ & - &\frac{16}{3\pi} (1-z)^{7/2} \ln(1-z) +\frac{2}{3}(1-z)^4\ln(1-z)+\mathcal{O}\left((1-z)^{9/2}\right)\,,\\
  s_5(z) & = &-\frac{32 (1-z)^3G(z) \Pi^{(1),v}(z)}{3 \pi z^2} +\frac{656}{81\pi^3 z}    \nonumber \\  & \nonumber   \,\,\mathop{\asymp}\limits^{z\to1} &
 \left[ -\frac{11}{8} + \ln(8) + \frac{3}{2 \pi^2}\left( -2 +7 \zeta_3 \right)  \right] (1-z)^{5/2}+(1-z)^{5/2} \ln(1-z)  \\ &- &\frac{2}{\pi} (1-z)^3 \ln(1-z)+\frac{1}{48}\left( -145 + 264 \ln(2)
     . + \frac{376 + 924 \zeta_3}{ \pi^2}\right) (1-z)^{7/2}+\frac{11}{6}(1-z)^{7/2} \ln(1-z)\nonumber  \\ & - & \frac{28}{3\pi}(1-z)^4 \ln(1-z)+\mathcal{O}\left((1-z)^{9/2}\right)\,, \\
  s_6(z) &=& -\frac{16 (1-z)^3 \Pi^{(1),v}(z)}{3 z^3}  +
      \frac{328}{81\pi^2 z^2}-\frac{6404 }{675 \pi^2 z}
   \nonumber \\  &    \,\,\mathop{\asymp}\limits^{z\to1} & (1-z)^3 \ln (1-z) -\frac{8}{\pi} (1-z)^{7/2} + \frac{7}{3} (1-z)^4 \ln (1-z)+\mathcal{O}\left((1-z)^{9/2}\right)\,,\\
   s_7(z) &=& -\frac{32 (1-z)^4 G(z) \Pi^{(1),v}(z)}{3 \pi z^3} +\frac{656}{81\pi^3 z^2}-\frac{131672 }{6075 \pi^3 z}
     \nonumber \\  &    \,\,\mathop{\asymp}\limits^{z\to1} &  \left[ -\frac{11}{8} + \ln(8) + \frac{3}{2 \pi^2}\left( -2 +7 \zeta_3 \right)  \right] (1-z)^{7/2} \nonumber \\ &+& (1-z)^{7/2}\ln (1-z) -\frac{2}{\pi} (1-z)^4 \ln (1-z)+\mathcal{O}\left((1-z)^{9/2}\right)\,,\\
s_8(z) &=&    -\frac{16 (1-z)^4 \Pi^{(1),v}(z)}{3 z^4}  +
      \frac{328}{81\pi^2 z^3}  -
     \frac{27412}{2025 \pi^2 z^2}
     +\frac{7773424 }{496125 \pi^2 z}
       \nonumber \\  &    \,\,\mathop{\asymp}\limits^{z\to1} & (1-z)^4 \ln (1-z) +\mathcal{O}\left((1-z)^{9/2}\right)\,,
 \label{eq:subs8}
\end{eqnarray}
where we have used the symbol $\asymp$ to denote that terms analytical
in $(1-z)$ have been dropped on the right-hand side, and we only use subtractions for the logarithmic terms,
hence no subtraction functions $s_1(z)$ and $s_3(z)$ are necessary.
We have used
\begin{equation}
 G(z) = \frac{1}{2z\sqrt{1-1/z}}\ln\left(\frac{\sqrt{1-1/z}-1}{\sqrt{1-1/z}+1}\right),
\end{equation}
and $\Pi^{(1),v}$ is the well-known two-loop correction to the vacuum
polarization~\cite{Kallen:1955fb} in the convention
of~\cite{Kiyo:2009gb}. The functions $s_i$ in eqs.~\eqref{eq:subs2} -- \eqref{eq:subs8} are
constant as $z\to0$ and only diverge logarithmically as $z\to\infty$.
\end{widetext}

\bibliography{biblioZZ}

\end{document}